\documentclass[12pt]{article}
\pdfoutput=1

\usepackage{cite}
\usepackage{booktabs}
\usepackage[english]{babel}
\usepackage{amsmath,amssymb,amsbsy,amstext, amsthm, simplewick}
\usepackage{hyperref}
\usepackage{graphicx}
\usepackage{amsfonts}
\usepackage{amssymb}
\usepackage[small]{caption}
\usepackage{upgreek}
\usepackage[svgnames,dvipsnames,x11names,table]{xcolor}
\usepackage{multirow}
\usepackage{geometry}
\usepackage[hang,flushmargin]{footmisc}
\usepackage{bm}
\usepackage{braket}
\usepackage{subcaption}
\usepackage{mathtools}
\usepackage{setspace}
\usepackage{cleveref}
\usepackage{comment}
\usepackage{scalerel}
\usepackage[normalem]{ulem}
\usepackage{slashed}
\usepackage{enumitem}
\usepackage{dsfont}
\usepackage{tikz}
\usetikzlibrary{decorations.markings}
\usetikzlibrary{shapes.misc}
\usepackage[export]{adjustbox}

\makeatletter
\g@addto@macro\bfseries{\boldmath}
\makeatother

\hypersetup{
    colorlinks=true,
    linkcolor={red!50!black},
    citecolor={blue!50!black},
    urlcolor={blue!80!black}
}

\usepackage{colortbl}

\makeatletter
\newlength{\apb@width}
\newcommand{\autoparbox}[2][c]{\settowidth{\apb@width}{#2}\parbox[#1]{\apb@width}{#2}}

\makeatother

\definecolor{lightgray}{gray}{0.9}

\usepackage[framemethod=default]{mdframed}
\newmdenv[skipabove=7pt,
skipbelow=7pt,
rightline=false,
leftline=false,
topline=false,
bottomline=false,
backgroundcolor=gray!10,
linecolor=gray,
innerleftmargin=5pt,
innerrightmargin=5pt,
innertopmargin=5pt,
innerbottommargin=5pt,
leftmargin=0cm,
rightmargin=0cm,
linewidth=4pt]{eBox}

\usepackage[most]{tcolorbox}
\tcbset{colback=white, colframe=black,
        highlight math style= {enhanced, 
            colframe=red,colback=red!10!white,boxsep=0pt}
        }
\definecolor{light-gray}{gray}{0.95}

\crefname{table}{Table}{Tables}
\crefname{equation}{Eq.}{Eqs.}
\crefname{appendix}{Appendix}{Appendices}
\crefname{section}{Section}{Sections}
\crefname{figure}{Figure}{Figures}

\numberwithin{equation}{section}

\def\beq{\begin{equation}}
\def\eeq{\end{equation}}

\def\bea{\begin{eqnarray}}
\def\eea{\end{eqnarray}}

\def\vp{\varphi_{+}}
\def\vm{\varphi_{-}}
\def\dvp{\dot{\varphi}_{+}}
\def\dvm{\dot{\varphi}_{-}}

\def\tp{\theta_{+}}
\def\tm{\theta_{-}}

\def\d{{\rm d}}

\def\beq{\begin{equation}}
\def\eeq{\end{equation}}
\def\bea{\begin{eqnarray}}
\def\eea{\end{eqnarray}}

\def\d{{\rm d}}

\def\O{{\cal O}}

\def\d{{\rm d}}

\def\k{{\vec{\scaleto{k}{7pt}}}}

\def\kp{{\!\!\vec{{\scaleto{\,\, k}{7pt}}^{\s\prime}}}}

\def\x{{\vec x}}

\def\t{\texttt{t}}
\def\V{{\cal V}}

\DeclareRobustCommand{\SkipTocEntry}[4]{}

\newcommand{\s}{\hspace{0.8pt}}

\definecolor{colorTC}{rgb}{.2,.7,.2}

\definecolor{amethyst}{rgb}{0.6, 0.4, 0.8}

\definecolor{acolor}{rgb}{0.4, 0.2, 0.4}

\definecolor{blue3}{RGB}{31, 119, 180}
\definecolor{red3}{RGB}{	214, 39, 40}
\definecolor{orange3}{RGB}{255, 127, 14}
\definecolor{green3}{RGB}{44, 160, 44}

\newcommand{\mb}[1]{{\mathbf{#1}}}
\newcommand{\ud}{\mathrm{d}}
\newcommand{\tFo}[4]{{}_2 F_1\!\left[\genfrac..{-1pt}{0}{\raisebox{-1pt}{$#1, \,\, #2$}}{\raisebox{1pt}{$#3$}} \, \bigg| \, #4\, \right]}

\newcommand{\tFt}[5]{{}_2 F_2\!\left[\genfrac..{-1pt}{0}{\raisebox{-1pt}{$#1, \,\, #2$}}{\raisebox{1pt}{$#3, \,\, #4$}} \, \bigg| \, #5\, \right]}
\newcommand{\comp}{\theta} 
\newcommand{\momcoeff}[2]{[G_{#1}]_{#2}}

\tikzstyle{intSty}=[draw=white, thick, line width=0.24mm]

\definecolor{pyBlue}{RGB}{31, 119, 180}
\definecolor{pyRed}{RGB}{214, 39, 40}
\definecolor{pyGreen}{RGB}{44, 160, 44}
\definecolor{pyBlue2}{RGB}{0, 111, 237}
\definecolor{pyRed2}{RGB}{224, 52, 36}

\begin{document}

\begin{titlepage}
\setcounter{page}{1} \baselineskip=15.5pt 
\thispagestyle{empty}
$\quad$
{\raggedleft CERN-TH-2025-154\par}
\vskip 60 pt
\begin{center}
{\fontsize{20.74}{18} \bf 
A Compact Story of Positivity
 in de Sitter
}
\end{center}

\vskip 20pt
\begin{center}
\noindent
{\fontsize{12}{18}\selectfont  Priyesh Chakraborty$^{\s 1}$, Timothy Cohen$^{\s 2,3,4}$, Daniel Green$^{\s 5}$, and Yiwen Huang$^{\s 5}$}
\end{center}

\begin{center}
\vskip 4pt
\textit{$^1$ {\small Department of Physics, Harvard University, Cambridge, MA 02138, USA}}
\vskip 4pt
\textit{ $^2${\small Theoretical Physics Department, CERN, 1211 Geneva, Switzerland}
}
\vskip 4pt
\textit{ $^3${\small Theoretical Particle Physics Laboratory, EPFL, 1015 Lausanne, Switzerland}
}
\vskip 4pt
\textit{ $^4${\small Institute for Fundamental Science, University of Oregon, Eugene, OR 97403, USA}
}
\vskip 4pt
\textit{ $^5${\small Department of Physics, University of California, San Diego,  La Jolla, CA 92093, USA}
}

\end{center}

\vspace{0.4cm}
 \begin{center}{\bf Abstract}
 \end{center}

\noindent
Recent developments have yielded significant progress towards systematically understanding loop corrections to de Sitter (dS) correlators. In close analogy with physics in Anti-de Sitter (AdS), large logarithms can result from loops that can be interpreted as corrections to the dimensions of operators. In contrast with AdS, these dimensions are not manifestly real.  This implies that the theoretical constraints on the associated correlators are less transparent, particularly in the presence of light scalars. In this paper, we revisit these issues by performing and comparing calculations using the spectral representation approach and the Soft de Sitter Effective Theory (SdSET). We review the general arguments that yield positivity constraints on dS correlators from both perspectives. Our particular focus will be on vertex operators for compact scalar fields, since this case introduces novel complications. We will explain how to resolve apparent disagreements between different techniques for calculating the anomalous dimensions for principal series fields coupled to these vertex operators. Along the way, we will offer new proofs of positivity of the anomalous dimensions, and explain why renormalization group flow associated with these anomalous dimensions in SdSET is the same as resumming bubble diagrams in the spectral representation.  

\end{titlepage}
\setcounter{page}{2}

\restoregeometry

\begin{spacing}{1.2}
\newpage
\setcounter{tocdepth}{2}
\tableofcontents
\end{spacing}

\setstretch{1.1}
\newpage

\section{Introduction}

Primordial non-Gaussianity offers the unique opportunity to probe the particle content and interactions of the inflationary era~\cite{Achucarro:2022qrl}. Single field inflation is highly constrained by symmetries~\cite{Salopek:1990jq,Weinberg:2003sw}, giving rise to an infinite family of single-field consistency conditions~\cite{Maldacena:2002vr,Creminelli:2004yq}. Additional fields can break these symmetries, which can yield unique signatures that can encode the mass, spin, and other properties of the additional degrees of freedom. These signatures of additional fields~\cite{Chen:2009zp,Achucarro:2010da,Baumann:2011nk,Noumi:2012vr,Chen:2012ge,Assassi:2012zq,Green:2013rd,Lee:2016vti,Chen:2016uwp,Gleyzes:2016tdh,Meerburg:2016zdz,Kumar:2019ebj,Jazayeri:2023xcj,Green:2023uyz,Cabass:2024wob,Sohn:2024xzd} is sometimes broadly described as {\it cosmological collider physics}~\cite{Arkani-Hamed:2015bza} and has motivated the {\it cosmological bootstrap} program~\cite{Baumann:2022jpr}, which offers a powerful way to understand the nature of these signals beyond their soft-limits~\cite{Assassi:2012zq,Mirbabayi:2015hva}. 

Understanding the space of possible inflationary signals is an important open problem. Theoretically, direct calculations at higher loop order are technically challenging and still poorly understood. A clear physical understanding of the constraints imposed on the correlators has proven extremely useful to both bootstrapping the results and in making sense of a variety of IR issues~\cite{Green:2022ovz}, but many fundamental questions remain open. Observationally, most cosmological collider signals require dedicated analyses and can be missed without a clear understanding of the space of signals. New insights into the structure of the calculation of cosmological correlators continues to inspire new ways to isolate new physics in cosmological data~\cite{Slosar:2008hx,Munchmeyer:2019wlh,Cabass:2022oap,Green:2023uyz,Cabass:2024wob,Philcox:2024jpd,Philcox:2025wts}.

Much of the structure of cosmological observables relevant for cosmic surveys already appears when studying the behavior of quantum fields in the simpler setting of a fixed de Sitter (dS) background.  Two principles that have driven recent developments in perturbative dS calculations are symmetries and positivity. A fixed dS background has a group $SO(4,1)$ of isometries~\cite{Spradlin:2001pw}. These generate Ward identifies for the late time correlators that are similar to a Euclidean Conformal Field Theory (CFT) in 3 dimensions~\cite{Strominger:2001pn,Maldacena:2011nz,Antoniadis:2011ib,Creminelli:2011mw}. Since the future surfaces are space-like, there is no manifest notation of unitarity associated with the operators appearing in these Ward identities. Yet, we know that the QFT in curved space must still obey some notions of positivity either from the spectral density~\cite{Bros:2010rku,Marolf:2010zp}, or from the wavefunction~\cite{Melville:2021lst,Goodhew:2021oqg,Meltzer:2021zin,Goodhew:2024eup}. The positivity of the two-point statistics of real fields has a long history in dS. For example, it is the basis for the Higuchi bound~\cite{Higuchi:1986py}, which limits that range of allowed masses. Positivity bounds are also observationally relevant in the form of the Suyama-Yamaguchi inequality~\cite{Suyama:2007bg,Smith:2011if} that relates the amplitudes of the three- and four-point functions from inflation. More recently, approaches relying on positivity of the spectral densities~\cite{Hogervorst:2021uvp,DiPietro:2021sjt,Loparco:2023rug}, and both the two-point~\cite{Cohen:2024anu} and four-point~\cite{Green:2023ids} statistics have been developed to constrain the spectrum of operators in dS.\footnote {All of these bounds crucially rely on the dS isometries, and therefore do not generalize to inflation~\cite{Green:2023ids}.} Finding the connection between these explicit results and the positivity bounds associated with scattering amplitudes~\cite{Pham:1985cr,Adams:2006sv,deRham:2022hpx} remains as one of the goals of the program to better understand cosmological correlators (see e.g.~\cite{Baumann:2015nta,Baumann:2019ghk,Grall:2021xxm,Creminelli:2022onn,Melville:2024ove,Pueyo:2024twm}).

Unfortunately, symmetries and positivity do not work together as seamlessly in dS as one might expect from AdS or flat space~\cite{Flauger:2022hie}. Dimensions of operators in dS can become complex~\cite{dsrep0,Bargmann:1946me,dsrep1,dsrep2,Sun:2021thf,Penedones:2023uqc}, ensuring that the statistics of individual scaling operators are not subject to positivity bounds. Still, real fields in dS should obey positivity constraints on their even point statistical correlators.  Surprisingly, these constraints cannot be satisfied using only the spectrum of operators determined at separated points and their Ward identities. Instead, positivity is maintained due to contact terms that naively violate the Ward identities~\cite{Cohen:2024anu}. In this sense, the interplay between positivity and symmetries is subtle: making one manifest can obscure the other.

In this paper, we will explore this interplay in the context of massless compact scalar fields. Massless fields in dS are special because they behave like dimension zero ($\Delta =0$) operators at long distances.  This implies that their two-point statistics grows logarithmically and that there is non-trivial  Renormalization Group (RG) mixing between any non-derivative composite operators. This behavior is similar a massless scalar in a two-dimensions~\cite{Ginsparg:1988ui}, which also suggests that working with operators given by the exponential of the field, so-called \emph{vertex operators}, can be used to define composite operators with fixed scaling dimensions. For compact scalars, we can further restrict to vertex operators that satisfy the periodicity constraint for the field. This complex behavior under RG flow turns out to be equivalent~\cite{Gorbenko:2019rza,Baumgart:2019clc,Mirbabayi:2019qtx,Cohen:2020php,Mirbabayi:2020vyt} to the formalism of stochastic inflation~\cite{Vilenkin:1983xq,Starobinsky:1986fx,Aryal:1987vn}. This connection between RG flow and stochastic inflation is manifest in the EFT approach of Soft de Sitter Effective Theory (SdSET)~\cite{Cohen:2020php,Cohen:2021fzf,Cohen:2021jbo}.

Of specific concern in this work are the anomalous dimensions for principal series (heavy) fields coupled to vertex operators of the (massless) compact scalars. Prior work found that the signs of anomalous dimensions are not bounded, and negative anomalous dimensions can arise at special values of the radius~\cite{Chakraborty:2023eoq}. This calculation used the Lorentzian inversion formula to determine the spectral density, which is known to yield the correct result when applied to non-vertex operator couplings~\cite{Chakraborty:2023qbp}. Yet, the result for compact scalars would appear to contradict the positivity constraints on the observable two-point~\cite{Cohen:2024anu} and four-point function~\cite{Green:2023ids}. This apparent disagreement demands explanation as it points to potential subtleties in the applications of both the EFT and the spectral density approach to computing dS correlators.

More generally, interactions with compact scalars offer a novel testing ground for our understanding of long distance effects in cosmology. Compact scalars display the same complexity under RG flow (or equivalently, stochastic inflation) as other lights field. However, their finite volume of field space also makes these theories directly solvable directly from the UV perspective, e.g.~using the path integral. On the other hand, understanding these results as a non-trivial RG flow also requires a treatment of vertex operators in SdSET that includes the full tower of operators. As a result, the compact scalar provides a solvable but still non-trivial case with which to stress test all perspectives.

The plan for the paper is as follows. In \cref{sec:setup}, we introduce the spectral density and SdSET techniques for calculating anomalous dimensions. We derive some identities showing that the results agree under some general assumptions. In \cref{sec:inversion}, we discuss the anomalous dimensions generated by vertex operators using the spectral density with the Lorentzian inversion formula and should how this must be evaluated to arrive at consistent results. In \cref{sec:sdset}, we explain how the same results arise in SdSET. We conclude in \cref{sec:conclusions}.

We will use the following conventions throughout. Correlators in momentum space are defined so that 
\beq
\langle \O(\k_1) \dots \O(\k_n) \rangle= \langle \O(\k_1) \dots \O(\k_n) \rangle' \, (2\pi)^3 \delta(\sum_i \k_i) \ .
\eeq
Spatial vectors are denoted $\k$ or $\x$ with lengths $k \equiv |\k|$ and $x = |\x|$. We will work in $(d+1)$-dimensions throughout, unless otherwise specified. The scaling dimensions of operators in the free theory will generally be defined as $\Delta$ and the dimension of the shadow operator associated with it are $\bar \Delta \equiv d- \Delta$. Interchanging these two solutions, via $\Delta  \to d-\Delta$, is referred to as a shadow transform (or shadow symmetry). In the interacting theory both acquire anomalous dimensions so that their respective scaling behavior is determined by $\Delta +\gamma$ and $\bar \Delta + \bar \gamma$, which generically breaks the shadow symmetry.

\section{Frameworks for Computing Anomalous Dimensions}\label{sec:setup}

Our primary goal in this paper is to demonstrate that anomalous dimensions in dS are consistent with positivity from several points of view. To set the stage, this section provides a review of how these anomalous dimensions arise generally within the frameworks of Lorentzian Inversion and Soft dS Effective Theory, before addressing the specific challenges associated with computing correlators for vertex operators of compact scalars.

\subsection{Positivity and Loop Corrections in the In-In Formalism}\label{sec:2.1}

Cosmic surveys provide a strong motivation for understand cosmological (in-in) correlators. Fluctuations of the metric generated during inflation can be predicted using quantum mechanical correlators in a fixed initial state. Current observations are consistent with purely adiabatic primordial fluctuations~\cite{Weinberg:2003sw}. These can be written purely in terms of scalar mode of the metric, $\zeta$~\cite{Bardeen:1980kt,Salopek:1990jq}. One common way to define $\zeta$ is by the working with ADM form of the metric\footnote{We can take the lapse and shift to be $N=1$ and $N^i=0$ respectively up ${\cal O}(\epsilon)$ slow-roll corrections~\cite{Cheung:2007st,Green:2024hbw}.}
\beq\label{eq:metric_zeta}
\d s^2 = -\d t^2 + a(t)^2 e^{2\zeta(\x,t)} \d\vec{x\s}^2 = a(\eta)^2 (-\d \eta^2 +e^{2\zeta(\x,\eta)} \d\vec{x\s}^2 )\, .
\eeq
 At late times, the fluctuations are effectively classical, and their statistical correlations can be determined from a wavefunction $\Psi[\zeta,t]$ via~\cite{Maldacena:2002vr}
\beq\label{eq:wavefunction_zeta}
\big\langle \zeta(\k_1,t) \dots \zeta(\k_n,t) \big\rangle = \int {\cal D}\zeta\, \zeta(\k_1) \dots \zeta(\k_n) \big|\Psi[\zeta,t]\big|^2\, . 
\eeq
These can be mapped onto observations of density fluctuations in our universe.

Motivated by these real observations, there is significant interest in understanding how to systematically determine in-in correlators for any quantum field theory on an FRW background. Quantum field theory on a fixed $d+1$ dimensional\footnote{We will generally be interested in $d=3$, but we will leave $d$ as a free parameter when possible.} dS background is a particularly compelling simplifying assumption, since the background is very similar to inflation but benefits from additional symmetries arising from the dS isometries. Concretely, we can learn a lot about cosmology from QFT correlators on
\beq\label{eq:metric}
\d s^2 = -\d t^2 + a(t)^2 \d\vec{x\s}^2 = a(\eta)^2 (-\d \eta^2 + \d\vec{x\s}^2 )\, , 
\eeq
where in pure dS, $a(t) = e^{Ht}$ or $a(\eta) =-1/(H\eta)$ with conformal time $\eta \in (-\infty, 0]$. On this background the in-in correlators satisfy the same Ward identities as a $d$ dimensional Euclidean CFT.

The observation that in-in correlators are determined by the wavefunction as in \eqref{eq:wavefunction_zeta} is true for any  operators ${\cal O}_i(\k\s)$ that can be written in terms of fields. For real operators, ${\cal O}_i^*(\k\s) ={\cal O}_i(-\k\s)$, which implies that 
\beq
\big\langle {\cal O}(\k\s) {\cal O}(-\k \s)\big\rangle'  \geq 0  \, .
\eeq
This basic positivity requirement, discussed in detail in~\cite{Green:2023ids,Cohen:2024anu}, has non-trivial consequences for cosmological collider-like signals. 

For (real) principal series scalars in dS, this apparently trivial positivity constraint is surprisingly powerful. Given such a field, $\phi$, with mass $m^2 > (dH/2)^2$,  the power spectrum in the long distance limit is given by
\begin{align}
\big\langle \phi(\k\s) \phi(-\k\s) \big\rangle' \to (- H \eta)^3\left( c   + c_\Delta (-k\eta)^{2i \nu}+ c_\Delta^* 
(-k\eta)^{-2i \nu} \right) \,,
\end{align}
where $\nu = \sqrt{\frac{m^2}{H^2} - \frac{d^2}{4}}$ and $d$ is the number of spatial dimensions. This result can be understood as a consequence of conformal invariance for an operator with a complex effective dimensions $\Delta_\phi = d/2  + i\nu$ and its complex conjugate with dimension $\Delta_\phi^*$. It is an important feature of the principal series fields that  
\begin{align}
\Delta_\phi^* = \bar \Delta_\phi \equiv d - \Delta_\phi\,,
\end{align}
where $\bar \Delta_\phi$ is the shadow dimension. The second and third terms do not have a definite sign, and therefore $c > 2 |c_\Delta|$ is required by positivity.

The situation becomes more complicated when we consider interacting theories. Perturbative corrections to the correlators are often described using the interaction picture to separate the time evolution of the free and interacting theory~\cite{Weinberg:2005vy}: 
\begin{align}
\langle {\cal Q}(t)\rangle=\left\langle\left[\bar{\text{T}} \exp \left(i \int_{-\infty_+}^t \d t\, H_{\rm int}(t) \right)\right] {\cal Q}^{\rm int}(t)\left[\text{T} \exp \left(-i \int_{-\infty_- }^t \d t\, H_{\rm int}(t)\right)\right]\right\rangle \ ,
\end{align}
where ${\cal Q}(t)$ is an operator defined at a single time $t$ but not necessarily local in spatial coordinates, ${\cal Q}^{\rm int}$ is the same operator defined in terms of interaction picture fields, and $\infty_\pm \equiv \infty (1 \pm i\epsilon)$. In the presence of these interactions, the scaling dimensions of operators in dS can acquire anomalous dimensions, much like their counterparts in flat space QFT.

For principal series fields, the presence of a real anomalous dimension $\gamma$ significantly changes the behavior of the correlation functions in a way that is not simply equivalent to a shift in the mass, since now $\Delta = \frac{d}{2}+\gamma + i\nu$. In the long distance description, the two point function for the interacting theory then takes the form
\beq\label{eq:phitwopoint}
\langle \phi(\k\s) \phi(-\k\s) \rangle' \to  (- H \eta)^3\left( c   + c_\Delta (-k\eta)^{2\gamma+2i \nu}+ c_\Delta^* 
(-k\eta)^{2\gamma-2i \nu} \right)\,.
\eeq
For $\gamma <0$ and $\nu \neq 0$, this two-point function is not positive in the $k|\eta| \to 0$ limit for any choice of $c$ and $c_\Delta$. This is a non-obvious property when one computes $\gamma$ in perturbation theory, despite the apparently trivial requirement that the power spectrum is positive.\footnote{The appearance of $c >0$ is required to maintain positivity when $\gamma > 0$. This is also non-variance as $c\neq 0$ is not allowed by the conformal Ward identities using dimension $\Delta$ and $\Delta^*$. An additional non-trivial consequence of positivity is the need for contact operators to generate this term~\cite{Cohen:2024anu}.}

\subsection{The Spectral Representation and Inversion Formulas}\label{sec:spectral_general}
Computing anomalous dimensions can be a cumbersome task.  In $k$-space, the relevant calculations can often yield complicated loop-integrals which can be difficult to evaluate and interpret, especially beyond one-loop. It is therefore useful, whenever possible, to express loop calculations as a sum over free-field exchanges in both dS \cite{Marolf:2010zp,Lu:2021wxu,Chakraborty:2023qbp} and AdS \cite{Fitzpatrick:2011hu, Giombi:2017hpr}. 
One fruitful approach results from first analytically continuing to Euclidean momentum space; the resulting formalism is intimately related K\"all\'en-Lehmann representation in dS. In this section, we will briefly review this theoretical technology, and will show how anomalous dimensions appear. For completeness, we recap the equivalent story in $(d+1)$-dimensional flat-space in \cref{app:2pcf_flat}.

\subsubsection{Preliminaries}
It will be convenient to work in position space, where all the two-point correlators $G(\vec{x},\vec{x}\s')$ are functions of the dS invariant embedding distance $\xi(\vec{x},\vec{x}\s')$. In the flat slicing this distance takes the form,
\begin{equation}
    \xi = 1- \frac{|\vec{x}-\vec{x}\s'|^2 - (\eta-\eta')^2+i\epsilon}{2\eta \eta'} \,,
    \label{eq:xiDef}
\end{equation}
where the $i \epsilon$ prescription is essential to ensure the appropriate time-ordering of the operators.

The two-point correlation function in position space can be written in the Watson-Sommerfeld form upon analytically continuing from the Euclidean sphere. This integral representation is \cite{Marolf:2010zp}
\begin{equation}\label{eq:watson_somm}
    G_\phi(\xi) = -\int_\mathcal{C} \frac{\ud J}{2\pi i} (d+2J)\s \momcoeff{\phi}{J}\s G(-J; \xi)\,,
\end{equation}
where the contour $\mathcal{C}$ runs parallel to the ${\rm Im}\,J$ axis, slightly to the left of the origin and $G(-J;\xi)$ is the free scalar two-point function with scaling dimension $\Delta=-J$. The term $\momcoeff{\phi}{J}$ is the Euclidean momentum coefficient and can be understood as the dS avatar of a momentum space propagator.\footnote{Note the distinction between $J$ and the spatial momentum $\vec k$ from Section~\ref{sec:2.1}} For a free field $\phi$, this is simply \cite{Marolf:2010zp,Chakraborty:2023qbp}
\begin{equation}
    \momcoeff{\phi}{J} = \frac{1}{J(J+d)+m_\phi^2} = \frac{1}{(J+\Delta_\phi)(J+\bar{\Delta}_\phi)}\,,
\end{equation}
where $\Delta_\phi= \frac{d}{2} + i\nu$ is the scaling dimension of $\phi$ and $\bar \Delta_\phi = d - \Delta_\phi$. Exactly as in flat-space, the poles of the momentum coefficient encode the infrared behavior of the correlator, i.e.~the limit $\xi \to -\infty$.\footnote{Strictly speaking, we mean late times $\eta,\eta' \to 0^{-}$ or large spatial distances $|\vec{x}-\vec{x}\s'|\to \infty$.} This can be derived by deforming the contour to pick up the pole closest to the contour. 

To develop some intuition for this approach, we first investigate the simplest example, which is free theory. In this case, the poles of $\momcoeff{\phi}{J}$ occur at $J=-\Delta_\phi$ and $J=-\bar{\Delta}_\phi$.  This pair of poles is shadow symmetric since $\Delta_\phi + \bar{\Delta}_\phi=d$. In the $\xi \to -\infty$ limit, the free propagator breaks down into a pair of shadow symmetric power laws
\begin{equation}
    G_\phi(\xi) \to \mathcal{A}(\Delta_\phi)\left(-\frac{\xi}{2}\right)^{-\Delta_\phi} + \mathcal{A}(\bar{\Delta}_\phi)\left(-\frac{\xi}{2}\right)^{-\bar{\Delta}_\phi} + \cdots\,,
    \end{equation}
where the prefactors are 
\begin{equation}
    \mathcal{A}(\Delta) \equiv \frac{1}{(4\pi)^{\tfrac{d+1}{2}}}\frac{\Gamma(\Delta)\Gamma(d-2\Delta)}{\Gamma(\tfrac{d+1}{2}-\Delta)} \, ,
\end{equation}
as a consequence of the normalization of free-fields in the Bunch-Davies state. Note that since $\bar \Delta_\phi = \Delta_\phi^*$, these expressions are not only shadow symmetric but also real.

When we turn on perturbative interactions for $\phi$, the two free-field poles shift slightly. Perturbative calculations yield corrections in the form of non-simple poles at $J=-\Delta_\phi$ and $J= -\bar \Delta_\phi$. Specifically, summing a set of bubble diagrams shifts the poles so that they appear at $J_* = -\Delta_\phi - \gamma_\phi$ and $\bar{J}_*= -\bar{\Delta}_\phi - \bar{\gamma}_\phi$, where $\gamma_\phi$ and $\bar{\gamma}_\phi$  can be interpreted as \textit{anomalous dimensions}. 
We enforce an on-shell renormalization scheme, where the anomalous dimensions are
real numbers:
the imaginary part of the anomalous dimension can always by absorbed by the mass $\nu_\phi$ \cite{Marolf:2010zp,Chakraborty:2023qbp}.
Moreover, shadow symmetry guarantees that $\bar{\gamma}_\phi = \gamma_\phi$. 
However, unlike flat-space, the residues at the poles $J_*$ and $\bar{J}_*$, which we denote as $\mathcal{R}$ and $\bar{\mathcal{R}}$ respectively, cannot both be set to their free-field values since the wavefunction counterterm, originating from the kinetic term in the action, is shadow symmetric. In the interacting theory the two poles of the propagator are not, therefore forbidding the residues of both to be simultaneously tuned to their free-field values. In our scheme we pick the wavefunction counterterm such that $\bar{\mathcal{R}}^*=\mathcal{R}=(-2i\nu_\phi + r)^{-1}$, where $r$ is some scheme-independent real number. We spell out the renormalization procedure in detail in \cref{app:anom_dims_euclidean}.

Therefore the momentum coefficient in an interacting theory on the principal series line can be determined by tracking the two shifted poles, which in turn fix the long distance ($\xi \to -\infty$) behavior of the interacting two-point function. Specifically we obtain 
\begin{align}
\label{eq:mom_coeff_interact}
 \momcoeff{\phi}{J} &\simeq \frac{1}{-2i\nu_\phi + r} \frac{1}{J+\Delta_\phi+\gamma_\phi}+\frac{1}{2i\nu_\phi+r} \frac{1}{J+\bar{\Delta}_\phi+\bar{\gamma}_\phi} \,.
\end{align}
We see then that the appearance of a non-trivial anomalous dimension is signaled by the fact that the sum of these two poles $\Delta_* + \bar{\Delta}_* = d + 2\gamma_\phi \neq d$. Up until this point, there appears to be no constraint on the allowed range of $\gamma_\phi$. We will see below that requiring positivity will lead to a bound on $\gamma_\phi$.

Finally, note that for a general operator $\Phi$, one can find additional poles, and they do not necessarily appear in a shadow symmetric set. Let us call the first pole we encounter for such an operator $J=-\Delta_1$. The Watson-Sommerfeld integral then yields
\begin{equation}\label{eq:spec_poles}
    G_\Phi(\xi) = {\rm Res}_{-\Delta_1}\left(\momcoeff{\Phi}{J}\right) (2\Delta_1-d) \mathcal{A}(\Delta_1) \left(-\frac{\xi}{2}\right)^{-\Delta_1} +\cdots\,.
\end{equation}
In this way, the poles of $\momcoeff{\Phi}{J}$ encode the dimensions scaling operators that appear in correlators of $\Phi$.

\subsubsection{The Spectral Representation}
Any unitary propagator in dS can be expressed as a spectral representation over free field states, i.e. a sum over principal and complementary series states. This follows the same arguments used in flat space via an application of the resolution of the identity. The K\"all\'en-Lehmann (KL) representation for an operator $\phi$ can be expressed as \cite{Bros:1990cu,Hogervorst:2021uvp}
\begin{equation}\label{eq:spectral_rep}
    G_\phi(\xi) = \int_{\tfrac{d}{2}-i\infty}^{\tfrac{d}{2}+i \infty} \frac{\ud \Delta}{2\pi i} \rho_\phi(\Delta) G(\Delta; \xi) + ({\rm complementary\,\, states})\,,
\end{equation}
where $G_\phi(\xi)$ and $G(\Delta;\xi)$ are defined in \eqref{eq:watson_somm}.
Evaluating this integral yields an explicit sum over principal series states; complementary series states appear as residues that are picked up as we analytically continue in some parameter which causes poles to cross the principal series line. The spectral function $\rho_\phi(\Delta)$ must satisfy two main properties.  First, it inherits shadow symmetry from the free theory propagator:
\begin{align}
\rho_\phi(\Delta)=\rho_\phi(d-\Delta)\,.
\end{align} 
Additionally, it must be positive when $\Delta$ lies on the principal series axis
\begin{align}
\rho_\phi(\Delta)\geq 0\,,
\end{align}
as a consequence of unitarity.

The KL representation can be derived using the Watson-Sommerfeld representation. It is convenient to change variables in \eqref{eq:watson_somm} to $\Delta=-J$ and to shift the contour onto the principal series line $\Delta \in \frac{d}{2}+ i \mathbb{R}$. The assumption that we do not encounter any poles when shifting this contour is equivalent to assuming that no complementary series states contribute to the KL representation. Once the contour is on the principal series line, we can exploit the shadow symmetry of the free propagator to isolate the shadow symmetric piece of the rest of the integrand, namely $(d-2\Delta)\momcoeff{\phi}{-\Delta}$, which derives \eqref{eq:spectral_rep}.  Furthermore, this calculation yields
\begin{equation}\label{eq:momcoefftospectral}
    \rho_\phi(\Delta) = \frac{1}{2}(d-2\Delta)\momcoeff{\phi}{-\Delta} + (\Delta \leftrightarrow \bar{\Delta})\,.
\end{equation}
Evaluating this result assuming a principal series free scalar, this implies 
\begin{align}
\rho^\text{free}_\phi(\tfrac{d}{2}+i \nu) = \tfrac{1}{2}\delta(\nu - \nu_\phi) + \tfrac{1}{2}\delta(\nu + \nu_\phi)\,, 
\end{align}
as one would expect. 
In general, we learn that
\begin{equation}
    \rho_\phi(\Delta) = 2\nu {\rm Im}\,\momcoeff{\phi}{-\frac{d}{2}-i \nu}\,,
\end{equation}
where we have assumed analyticity for the momentum coefficients $\momcoeff{\phi}{J}^{*} = \momcoeff{\phi}{J^*}$. 

Next, we can apply the same logic to an interacting theory. Assuming the analyticity of the momentum coefficients still holds, \eqref{eq:mom_coeff_interact} then implies
\begin{equation}
    \rho_\phi\big(\tfrac{d}{2}+i \nu\big) \simeq \frac{1}{r^2+4\nu_\phi^2}\frac{16 \gamma_\phi  \nu ^2}{(\nu^2 -\nu_\phi^2)^2 + 2\gamma_\phi^2(\nu^2 + \nu_\phi^2) + \gamma_\phi^4}\,,
\end{equation}
where we have taken $\nu \simeq \pm \nu_\phi$.
This is the de Sitter avatar of a Breit-Wigner distribution.\footnote{If we restore factors of Hubble and approximate $\nu \simeq \mu/H$, $\nu_\phi \simeq m_\phi/H$ and $\gamma \simeq \Gamma/H$, and subsequently take the $H \to 0$ limit we recover the usual flat-space distribution $\rho(\mu)/(4 \mu H) \simeq \mu_\phi \Gamma/[(\mu^2-m_\phi^2)^2 + 4 m_\phi^2 \Gamma^2]$. Note that we have further assumed $\mu \simeq m_\phi$, as usual, and the factor of $1/(4\pi \mu H)$ is present to match the normalization convention in flat-space.} Positivity of $\gamma_\phi$ is then essential to ensure the positivity of the quantum corrected spectral density.  
As in flat-space, the anomalous dimension may be interpreted as a decay-width and the principal series field $\phi$ as a resonance.\footnote{This analogy is fraught, however, since even in the free theory the scalar $\phi$ decays due to Hubble dilution.} 

We can make the statement of positivity even more precise by explicitly coupling $\phi$ to some scalar operator $\Phi$, which could be a fundamental field or a composite operator.\footnote{Derivative interactions can also be recast into this form via integration by parts.} Let us assume that the interaction is
\begin{equation}\label{eq:general_interaction}
    S_{\rm int} = \int_x g\s \phi\s \Phi\,.
\end{equation}
The one-loop anomalous dimension for $\phi$ in this theory is (we review this calculation in \cref{app:anom_dims_euclidean})
\begin{equation}\label{eq:anomalous_dim_from_spec}
    \gamma_\phi = \frac{g^2}{2\nu_\phi} {\rm Im}( \momcoeff{\Phi}{-\Delta_\phi}) = \frac{g^2}{4\nu_\phi^2} \rho_\Phi(\Delta_\phi)\,,
\end{equation}
where we have leveraged the relationship between the momentum coefficient and the spectral density written above. We can therefore explicitly see that $\gamma_\phi \geq 0$ comes from the assumption that $\Phi$ is an operator in a unitary theory.\footnote{The same point was also recently made in \cite{Loparco:2025azm}.}

In passing, we note that this expression reduces to the one familiar to us from flat-space. To make this connection we can explicitly take the $\nu \to \infty$ limit of the dS spectral density to find
\begin{equation}
    \int \frac{\ud \nu}{2\pi} \rho_{\Phi}(\Delta) \simeq \int \ud \mu^2 \frac{\rho_{\Phi}(\Delta)}{4\pi \mu}\,,
\end{equation}
where we have approximated $\nu \simeq \mu$. We can therefore identify $\rho_{\Phi}^{\rm flat}(\mu^2)\simeq \rho_{\Phi}(\tfrac{3}{2}+i \mu)/(4\pi \mu)$ and
\begin{equation}
   \gamma_{\phi} \simeq g^2  \frac{\pi}{m_\phi} \rho_{\Phi}^{\rm flat}(m_\phi^2)\,,
\end{equation}
which is exactly the usual flat-space decay width expressed in terms of the flat-space spectral density. Since $\phi$ is principal series, $\rho_\Phi(\Delta_\phi)\propto |\langle 0|\Phi|\Delta_\phi \rangle|^2$, i.e.~the spectral density measures the leakage of probability from $\Phi$ into the $\phi$ sector. Concretely, this anomalous dimension arises from evaluating the spectral density of $\Phi$ at $\Delta =\Delta_\phi$; this is analogous to the energy conservation constraint for a flat-space decay process. For unitary CFTs, this connection can be made precise via the state operator correspondence, but there is no such correspondence in dS.

In situations where the spectral density is calculable, \eqref{eq:anomalous_dim_from_spec} provides an explicit formula to compute $\gamma_\phi$. Additionally, if we know the position space two-point function $G_\Phi(\xi)$, the spectral density $\rho_\Phi(\Delta)$ can be determined using so-called inversion formulae, which we review next.

\subsubsection{Inversion Formulae}
The spectral density can be computed by Wick rotating to Euclidean signature. We will do so using results derived on the flat-slicing of Euclidean Anti de Sitter (EAdS), on which $\xi \in (-\infty, -1)$. Points in dS which have distance $\xi<-1$ are causally disconnected and have superhorizon separations. Following \cite{Loparco:2023rug}, we leverage the orthonormality of the EAdS propagators to derive 
\begin{equation}\label{eq:eads_inversion}
    \rho_\Phi(\Delta) = \mathcal{N}_{\rm E}(\Delta) \int_{-\infty}^{-1} \ud \xi (\xi^2-1)^{\frac{d-1}{2}}\tFo{\Delta}{\bar{\Delta}}{\frac{d+1}{2}}{\frac{1+\xi}{2}} G_\Phi(\xi)\,,
\end{equation}
where ${}_2F_1[a,b;c;z]$ is the Gauss hypergeometric function and 
\begin{equation}
    \mathcal{N}_{\rm E}(\Delta) \equiv \frac{4\pi^{\tfrac{d+3}{2}}}{\Gamma(\Delta-\tfrac{d}{2})\Gamma(\tfrac{d}{2}-\Delta) \Gamma(\tfrac{d+1}{2})}
\end{equation}
is the the EAdS normalization.

\begin{figure}[t!]
    \centering
    \includegraphics[width=0.55\linewidth]{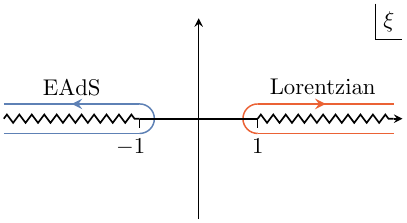}
    \caption{We illustrate the contours corresponding to the two inversion integrals for the spectral density. The orange contour represents the Lorentzian integral: this picks up the physical discontinuity of the propagator on time-like separated points, which admits $\xi>1$. The blue contour  represents Euclidean superhorizon separations, and are thus accessible in Euclidean AdS. On the Euclidean side, the branch cut of the integration kernel is exploited instead.}
    \label{fig:inversion_contours}
\end{figure}

The spectral density $\rho_\Phi(\Delta)$ can also be determined using Lorentzian, or timelike separated points \cite{Hogervorst:2021uvp}. This can be obtained by recognizing that the EAdS kernel can be expressed as a discontinuity over the range $\xi\in(-\infty,-1)$
\begin{equation}
    \tFo{\Delta}{\bar{\Delta}}{\frac{d+1}{2}}{\frac{1+\xi}{2}} \propto {\rm disc}\,\tFo{1-\Delta}{1-\bar{\Delta}}{\frac{3-d}{2}}{\frac{1-\xi}{2}}\,,
\end{equation}
which allows us to deform the integration contour over the complex $\xi$ plane, which is represented in blue in Figure~\ref{fig:inversion_contours}.\footnote{Greens functions in the Bunch-Davies state are analytic everywhere on the complex $\xi$ plane except for the time-like branch cut along $\xi \in (1,\infty)$ \cite{Spradlin:2001pw}. This not true for propagators in the so-called alpha vacua which admit a branch cut along $\xi \in (-\infty,-1)$. We are only concerned with Bunch-Davies correlators here so we are free to deform the contour in this manner.}  Placing this contour onto the timelike branch cut of the propagator $G_\Phi(\xi)$ along $\xi\in(1,\infty)$ yields
\begin{subequations}
\label{eq:lorentz_inv}    
\begin{equation}\label{eq:lorentz_inv_int}
    \rho_{\Phi}(\Delta) = \mathcal{N}_{\rm L}(\Delta) \int_1^{\infty} \ud \xi\, \tFo{1-\Delta}{1-\bar{\Delta}}{\tfrac{3-d}{2}}{\frac{1-\xi}{2}} {\rm disc}\,G_{\Phi}(\xi)\,,
\end{equation}
where the normalization is
\begin{equation}
    \mathcal{N}_{\rm L}(\Delta) \equiv -i \frac{(4 \pi )^{\frac{d+1}{2}}}{2\Gamma \left(\frac{3-d}{2} \right)} \frac{\Gamma (1-\Delta ) \Gamma (1-\bar{\Delta})}{\Gamma (\Delta -\tfrac{d}{2} ) \Gamma (\bar{\Delta}-\tfrac{d}{2})} \, .
    \label{eq:lorentz_inv_norm}
\end{equation}
\end{subequations}

We illustrate the contour choices corresponding to the two inversion formulae in Figure~\ref{fig:inversion_contours}. Note that the Lorentzian inversion integral sees coincident points ($\xi=1$). Of course, propagators are usually very singular at short distances, so  divergences can arise when evaluating this expression. As such, the Lorentzian inversion formula is technically defined via dimensional regularization.  One deforms the $\xi$-integration contour along the Lorentzian cut in $d$  dimensions.  For sufficiently small $d$, there are no coincident singularities at the branch point of the propagator so the contour can be evaluated.  One subsequently analytically continues $d \to 3$, picking up any potential poles in $d$ along the way.\footnote{We note that it is also possible to obtain a Lorentzian inversion formula for the momentum coefficient $\momcoeff{\phi}{J}$ by exploiting the Froissart-Gribov trick on the Euclidean sphere \cite{Hogervorst:2021uvp,Chakraborty:2023qbp}. It therefore suffers from the same short-distance subtleties as the equivalent inversion formula for the spectral density, but formally contains equivalent information.} 

As discussed around \eqref{eq:spec_poles}, the spectral density encodes the scaling operators that appear in dS correlators. In principle, the above expressions should have an interpretation directly in terms of operators defined a long distances. To make this connection, we will need the language of SdSET to organize the scaling operators of the theory.

\subsection{Soft de Sitter Effective Theory and Anomalous Dimensions}\label{sec:SdSET}
Effective Field Theory (EFT) is a powerful tool for organizing quantum field theories according to symmetries and RG flow. An EFT perspective on the origin of anomalous dimensions in dS would naturally be expected to provide some additional insights into the origins of positivity.

A natural EFT emerges when the observables of the theory are evaluated on super-horizon scales $k/a \ll H$.  This EFT is the Soft dS Effective Theory (SdSET), which organizes the scaling behavior for the dynamical modes of the theory in the long wavelength (soft) limit according to their time evolution. For example, given real scalar $\phi$ with mass $m$, solutions to the Klein-Gordon equation in dS motivate writing the field operator for the free theory as a mode expansion
\beq
\phi(\x,t) = H\! \left( [a(t)H]^{-\Delta} \varphi_+(\x,t) + [a(t)H]^{-\bar \Delta} \varphi_- (\x,t)\right)\, ,
\eeq
where $\Delta = \frac{d}{2} + i\nu$ for the principal series and $\bar \Delta = d-\Delta$. In the regime $k \ll a H$, the fields $\varphi_+$ and $\varphi_-$ behave as scaling operators of dimension $\Delta$ and $\bar \Delta$ respectively.

The goal of the SdSET is to make the long distance behavior of the theory apparent as a direct consequence of symmetries and power counting. The energy scale $a(t) H =\Lambda(t)$ is interpreted as the UV cutoff of the EFT and therefore coefficients of interactions involve appropriate powers of $\Lambda(t)$ that render the action dimensionless. As such, the role of each operator in the action can be determined by power counting in terms of $\lambda \sim k/(aH) \ll 1$, which provides the generalization of dimensional analysis for scaling operators in more typical EFTs. For a principal series field, the leading action is given by
\beq
S_2 =\int \d^d x\, \d\t \, i\nu \left(\dvp \vp - \dvm \vp \right)\,,
\eeq
where $\t = t H$ is dimensionless time and $\dot X \equiv \frac{\partial}{\partial \t} X$. Any mass terms can be removed by a redefinition of the fields and $\nu$~\cite{Cohen:2020php, Cohen:2024anu}. This action describes the super-horizon evolution of the modes starting from stochastic initial conditions that are determined by matching\footnote{In SdSET, any time-independent contributions to correlators can be absorbed into the initial conditions, and therefore can also match both vacuum fluctuations and open quantum systems~\cite{Salcedo:2024smn,Green:2024cmx,Colas:2025ind}. For vacuum fluctuations in $\lambda \phi^4$, the matching to SdSET was explicitly verified to order $\lambda^2$ in~\cite{Cohen:2021fzf} and proven to all orders in~\cite{Green:2025hmo}. Yet, as we will see, the field content of SdSET does not to include contributions for both light and heavy fields.}

In FRW slicing, the metric \eqref{eq:metric} is invariant under the scaling $\x \to e^{-\rho} \x$ and $a(t) \to e^{\rho} a(t)$, for any constant $\rho$. This rescaling is just a redundancy in the relationship between comoving and physical coordinates. Under this transformation, $\d\t$ is scaleless so the measure transforms as $e^{-d \rho}$.  Therefore, power counting and symmetries dictates that self-interactions take the form 
\beq\label{eq:sdset_int}
S_{\rm int} \supset \int \d^d x\, \d\t\, g_{n,m} [a(t)H]^{d-(n+m)\frac{d}{2} -i (n-m)\nu} \vp^n \vm^m\,,
\eeq
where the $g_{n,m}$ are dimensionless coupling constants.
Since $n+m \geq 3$, this operator is always  ``irrelevant" (for $d>0$) in the sense that the corrections are suppressed by (real) powers of $g_{n,m}\times k /\Lambda(t)$. Operators with spatial derivatives are additionally suppressed.

Within SdSET, generating interesting long distance behavior for massive fields requires coupling to additional operators~\cite{Cohen:2024anu}. 
Specifically, suppose we have scaling\footnote{We will use the notation $\O$ for operators with well-defined scaling dimensions, as opposed to $\Phi \to \sum_n \O_n$ which does not a behave as a single power law at long distances.} operators $\O$ and $\bar \O$ with dimensions $\Delta_\O$ and $\bar \Delta_\O$ respectively. With these dimensions, the operators can have a dS invariant commutator
\beq
\big\langle [\bar \O(\k,\t), \O(\kp,\t')]\big\rangle = \theta(\t'-\t) C(\Delta_\O) (2\pi)^d \delta(\k+\kp)\,,
\eeq
where $C(\Delta_\mathcal{O})$ can be derived from matching to the UV theory. Specifically, for interaction of the form \eqref{eq:general_interaction}, $\Phi \supset \O, {\bar \O}$ and $C(\Delta_
\mathcal{O})$ \footnote{The matching coefficient here is twice what's presented in \cite{Cohen:2024anu} because we have included both terms in the commutator.} is given by 
\begin{align}
C(\Delta_\mathcal{O}) = \frac{2 i}{d/2-\Delta_\mathcal{O}}\text{Res}_{\Delta = \Delta_\mathcal{O}}\rho_\Phi(\Delta)\,.
\label{eq:CDeltaRes}
\end{align}
If we couple two such operators to our principal series fields via
\begin{align}\label{eq:Ocoupling}
S_{\rm int} \supset \int \d^d x \s \d\t \,  g \bigg[ &\vp \Big([aH]^{d/2 -i \nu - \Delta_\O} \O +[aH]^{d/2 -i \nu - \bar \Delta_\O} \bar \O \Big)\notag\\[3pt]
 &+\vm \Big([aH]^{d/2 +i \nu - \Delta_\O} \O +[aH]^{d/2 +i \nu - \bar \Delta_\O} \bar \O \Big)\bigg] \,,
\end{align}
where $g$ is the same coupling as the UV by definition, as we have absorbed any non-trivial matching into the $C(\Delta)$ and the power spectrum of $\O$. We notice that the coupling to $\bar \O$ is relevant and can contribute at long distances. However, for stable\footnote{If the operator $\bar \O$ had its own fluctuations, when $\bar \Delta_\O < 0$ the fluctuations would grow at large separations (in time or space) and therefore would signal an unstable vacuum. This is the case of fields with $m^2 < 0$ where one finds $\Delta_\O < 0$.} theories in dS, $\bar \O$ does not have dynamics of its own. Concretely, we require that
\beq
\langle \bar \O(\k\s) \bar \O(\kp) \rangle  = 0 \, ,
\eeq
even when $C(\Delta_\O) \neq 0$. In this sense, $\bar \O$ only exists to encode contact terms that appear at long distances and are required to match the UV theory predictions. These contact terms are required to have $c > 0$ in \eqref{eq:phitwopoint} to obtain results that are consistent with positivity~\cite{Cohen:2024anu}. Note that this is only possible for a unitary theory in dS because there is no state operator correspondence. Specifically, the vanishing of the two point function of $\bar \O$ does not imply the presence of any null or negative norm states. Nevertheless, it is the difference in the power spectrum for $\O$ compared to $\bar\O$ that is 
dynamically responsible for breaking the shadow symmetry of the action in \eqref{eq:Ocoupling}. This is crucial for understanding how $\vp$ and $\vm$ acquire anomalous dimensions that break the shadow symmetry.

The presence of $\O$ and $\bar \O$ together generate a log-correction to the $\vp$ power spectrum~\cite{Cohen:2024anu}
\beq\label{eq:SdSETlog}
\langle \vp(\k\s) \vp(\kp) \rangle' = k^{2 i \nu} \left(1+g^2 \frac{\log (-k \eta)}{2 \nu}\frac{C(\Delta_\mathcal{O})}{\Delta_\mathcal{O}-\Delta_\phi}  \right)\, .
\eeq
The appearance of these time-dependence logs signals the need for dynamical RG~\cite{Green:2020txs}, where the coefficient of the log can be interpreted as an anomalous dimension for $\varphi_\pm$. In SdSET, the UV cutoff $\Lambda(t) = a(t) H$ is time-dependent and therefore re-summation via conventional RG determines time evolution, so-called ``dynamical RG.'' When multiple operators have similar dimensions, this evolution is more complicated and one finds a matrix of anomalous dimensions as usual in EFTs. 

The central appeal of SdSET is that the action \eqref{eq:sdset_int} makes it manifest when to expect interesting corrections to the long distance physics. Most interactions are irrelevant in SdSET, which explains why so much of the physics of dS space is determined at horizon crossing, $k=aH$. However, the anomalous dimensions generated for $\phi$ by $\O$ with $\Delta_\O \gg 1$ would appear to contradict this expectation. The need for the contact term between $\O$ and $\bar \O$ partially explains this because the coupling to $\bar \O$ is relevant and required to generate the anomalous dimension. However, because $\bar \O$ is itself not a dynamical field, this implies that it is not a relevant deformation of the theory: we only get a marginal contribution from the combined impact of $\O$ and $\bar \O$. This observation is crucial for understand how these results match those of the spectral decomposition, as we will see in this paper. 

\section{Anomalous Dimensions and Bubble Diagrams}
\label{sec:connecting}

One of the challenges for understanding physics in dS has been the lack of reliable physical intuition and non-perturbative principles on which perturbative results can be anchored. Both the spectral decomposition and SdSET aim to remedy this situation by appealing to flat space \& (unitary) CFT, and EFT \& superhorizon intuition respectively. Unfortunately, at times these different perspectives give rise to somewhat contradictory intuition. Before exploring new vistas (namely correlation functions of compact scalar vertex operators), we would like to understand how to connect these two points of view within the context of the calculation of anomalous dimensions for general scalar interactions.

At a conceptual level, from an EFT (or even CFT) point of view, 
it is surprising the anomalous dimension $\gamma_\phi$ in \eqref{eq:mom_coeff_interact} can be derived by summing only bubble diagrams. In SdSET, we will resum the leading logs using RG to find this anomalous dimension. Moreover, it is \emph{not} the case that in a typical QFT that RG flow is equivalent to summing a geometric series. A separate point of view is that the intuition behind these anomalous dimensions can be derived by analogy with the flat space decay width, as we have mentioned above.  The origin of the Breit-Wigner propagator in flat space can be traced to summing a Dyson series, which at one-loop is a geometric series of bubble diagrams (for a theory with cubic interactions.)  In the case of (A)dS, the statement that the anomalous dimensions can be computed by simply summing bubble diagrams instead holds to all orders (at least in this case).  In dS, we will see that this can be understood from the power counting of the SdSET.

The resolution to this confusion resides in SdSET and, specifically, the unique way in which the anomalous dimensions arise for principal series fields. The purpose of SdSET is that power counting makes it transparent when and how non-trivial long distance physics can arise. From \eqref{eq:sdset_int}, we saw that interactions for massive fields should be trivial in the IR since the interactions of the SdSET fields are always irrelevant by power counting. However, the presence of the shadow operators $\bar {\cal O}$ 
allows these fields to influence the
IR behavior because their dimensions are effectively negative, so that it can generate relevant operators. However, this field is non-dynamical, meaning that these apparently relevant interactions have no direct impact and instead only contribute to the theory via commutators with the long distance $\O$ operators. These contact interactions always take the form of tree-level diagrams and the logs can thus be resummed directly. This is not true for light fields in SdSET, where the long distance theory is genuinely interacting and a proper RG treatment gives rise to Stochastic Inflation, and corrections thereto~\cite{Cohen:2020php, Cohen:2021fzf}. Therefore it is a special feature of the non-dynamical nature of the $\bar \O$ operators that gives rise to this equivalence of RG and bubble diagrams.

Having understood the origin of the anomalous dimension on a conceptual level, we encounter an apparent tension on a technical level. In SdSET, anomalous dimensions arise from a sum over the entire tower of local operators representing the loop diagram. For example, in~\cite{Cohen:2024anu} the logarithmic term involves the sum over the scaling operators contained in $\phi^2$, namely
\beq 
\langle \vp(\k\s) \vp(\kp) \rangle' = k^{2 i \nu} \left(1+ g^2 \frac{\log (-k \eta)}{2 \nu_\phi}\sum_{j=0,\pm 1} \sum_{n=0}^\infty \frac{C(\Delta_{j,n})}{\Delta_{j,n}-\Delta_\phi}  \right)\, .
\eeq
where $\Delta_{j,n} = d + 2i j \nu_\phi +2 n$. The sum over $j$ is a consequence of the fact that the scaling operators must appear in complex conjugate pairs for any real operator $\phi^2$.
The coefficient of this log is twice the anomalous dimension of one of these operators, so that
\beq \label{eq:anom_dim_from_sum}
\gamma_\phi + i \delta \nu_\phi = \frac{g^2}{4\nu_\phi}\sum_{j=0,\pm 1} \sum_{n=0}^\infty \frac{C(\Delta_{j,n})}{\Delta_{j,n}-\Delta_\phi} \ .
\eeq
We have included a possible correction to the imaginary component, $\delta \nu_\phi$, which we can set to zero by a shift of the mass. At face value, this formula is substantially more complicated than \eqref{eq:anomalous_dim_from_spec}: $    \gamma_\phi = \frac{g^2}{4\nu_\phi^2} \rho_\Phi(\Delta_\phi)$. In addition, the anomalous dimension in SdSET is an infinite sum that is not even necessarily convergent. 

In order to reproduce \eqref{eq:anomalous_dim_from_spec}, we recall \eqref{eq:CDeltaRes}, which tells us that the coefficient of the contact terms are just the residues of the spectral density:
\beq
C(\Delta_\mathcal{O}) = \frac{2 i}{d/2-\Delta_\mathcal{O}}\text{Res}_{\Delta = \Delta_\mathcal{O}}\rho_\Phi(\Delta)\, .
\eeq 
We recall that the list of operators that appear in SdSET is precisely the list of poles in the spectral density, so we can rewrite the sum 
\beq
\gamma_\phi + i \delta \nu_\phi  = \frac{g^2}{2\nu_\phi} \oint \frac{\d\Delta}{2\pi i}  \frac{1}{\Delta-\Delta_\phi} \frac{i}{\Delta-d/2} \rho_\Phi(\Delta)\, .
\eeq
 where the contour picks up only the poles in $\rho_\Phi$ which correspond to the individual operators in spectrum.

This expression is identical to that of the spectral density, as long as we can ignore the contour at infinity. If this is the case, we find
\beq\label{eq:anom_contour}
 \oint \frac{\d\Delta}{2\pi i}  \frac{1}{\Delta-\Delta_\phi} \frac{i}{\Delta-d/2} \rho_\Phi(\Delta) = \int^{i\infty +\frac{3}{2}}_{-i\infty +\frac{3}{2}}\frac{\d\Delta}{2\pi i}  \frac{1}{\Delta-\Delta_\phi} \frac{i}{\Delta-d/2} \rho_\Phi(\Delta) \, .
\eeq  
Importantly, note that there are two poles on the principal series line at $\Delta  = \Delta_\phi = \frac{d}{2} + i\nu_\phi$ and $\Delta= \frac{d}{2}$. The latter, in fact, do not contribute to the contour integral since the spectral density vanishes at $\Delta=\frac{d}{2}$.\footnote{The free Green's function is singular for $\Delta=d/2$ due to the degeneracy $\Delta=\bar{\Delta}$. In the KL representation, it is therefore crucial for the spectral density $\rho_\Phi(\Delta)$ to vanish at this point. In fact, the free propagator is singular at all $\Delta= d/2+n$, for positive integer $n$. Thus in order to ensure we do not pick up any spurious contributions from the KL representation in the infrared, the spectral density must vanish for all of these values of the scaling dimension (see e.g. \cite{Hogervorst:2021uvp}).}
Taking the real part to isolate $\gamma_\phi$ (being careful to deform the contour around the pole at $\Delta_\phi$), one finds
\beq
\gamma_\phi =  \frac{g^2}{4\nu_\phi^2} \rho_\Phi(\Delta_\phi)  \, ,
\eeq
in exact agreement with the spectral representation result.

The fact that these anomalous dimensions arise from resumming the tree-level commutator of $\O$ and $\bar \O$ implies that the contribution is equivalent to the anomalous dimensions that arise via mass mixing, except that the dimensions of $\O$ are not confined to the principal series. The way we deform the contour in \eqref{eq:anom_contour} makes this manifest: the sum becomes an integral over principal series fields that mix with $\phi$. It is only through the evaluation of this contour integral, and specifically the mixing at $\Delta =\Delta_\phi$ where the fields are degenerate (and therefore cannot be trivially diagonalized), that we find a non-trivial real contribution to the anomalous dimension.

We should stress that the assumption that $\rho_\Phi(\Delta)$ vanishes along the contour at infinity is not, generally speaking, a valid assumption. It is typically the case that the density of states grows as a power-law at larger $\Delta$ and this contribution does not vanish. This manifests itself in the SdSET formulas, such as \eqref{eq:anom_dim_from_sum}, as a non-convergent sum. This behavior is due to familiar UV divergences in the flat space limit, and therefore signal that we must renormalize the theory in flat space. We emphasize that this is distinct from the RG in SdSET which resumms IR effects in the full UV theory, and leads to dynamical RG.  Physical regulators can be introduced that ensure that the contour at infinity does not directly contribute to $\gamma_\phi$, as it can only alter the couplings of the UV Lagrangian. We will provide one such example of a regulator for SdSET in \cref{sec:SdSET}, and a related but different regulator for the Euclidean sphere in Appendix~\ref{app:dispersive}. A more general treatment of the divergences at large $\Delta$ should follow from an analogous study performed for AdS~\cite{Fitzpatrick:2010zm} but will be left to future work.

\section{Lorentzian Inversion for Compact Scalars }\label{sec:inversion}
In this section, will calculate the anomalous dimension of a principal series scalar $\phi$ when coupled to a massless compact scalar field $\comp$ via a vertex operator $\mathcal{V}$. The interaction we assume is of the form \eqref{eq:general_interaction}, with coupling $g_\theta$, and we defer more details to a later part in this section, noting only the final result for convenience. Following our discussion in \cref{sec:spectral_general}, the anomalous dimension of $\phi$ can be straightforwardly obtained in terms of the spectral density of the vertex operator using \eqref{eq:anomalous_dim_from_spec}, 
\begin{equation}\label{eq:phi_anomlous_dim}
    \gamma_\phi = \frac{g_\comp^2}{4\nu_\phi^2} \rho_{\mathcal{V}}(\Delta_\phi)\, .
\end{equation}
It remains then to determine the spectral density $\rho_\mathcal{V}(\Delta)$. First, we will review how compact scalars and vertex operators behave in dS, and we will then couple $\phi$ to $\comp$ via such a vertex operator and determine the resulting anomalous dimension.

\subsection{Compact Scalars in dS}
We will begin with a free compact scalar, defined by the action $S = \int \tfrac{1}{2}f_\comp^2(\partial \comp)^2$,
where we are choosing to work with a dimensionless $\comp$ and $f_\comp$ is its decay constant. We are defining $\comp$ to be periodic such that there is a gauge\footnote{The periodicity of the compact scalar is a redundancy in the field description
rather than a mere symmetry of the potential, and therefore it has the interpretation of a gauge symmetry, with the consequence that physical operators must be gauge invariant \cite{Reece:2025thc}. As emphasized in \cite{Chakraborty:2023eoq}, this distinction is crucial as it enables the equilibrium distribution of the massless compact scalar to exist, and therefore for the theory to be well defined in dS.} constraint (equivalence), $\comp(x) \sim \comp(x) + 2\pi$. As a result, we are required to work with only the appropriate gauge invariant operators, which are $\partial_\mu \comp$ and $e^{i p \comp}$, where $p$ is an integer. We will only work with the latter, i.e.~vertex operators. To be precise, $e^{ip \comp}$ is highly composite so it is necessary to work with the normal ordered operator, 
\begin{align}
\mathcal{V}_p(x) \equiv\, :e^{i p \comp(x)}: \ .
\end{align}
As we will see, this can be interpreted as one choice of scheme for the renormalized vertex operator.

We will now discuss correlators of vertex operators for the free massless compact scalar, summarizing the results of \cite{Chakraborty:2023eoq}. The position space correlators of vertex operators are straightforward to determine in the free theory, by simply completing the square in the path integral. The resulting two-point function is given by 
\begin{equation}
    \big\langle e^{i p \comp(x)} e^{i q \comp(y)}\big\rangle = \delta_{p+q} e^{-2 p^2 \frac{H^2}{f^2}G_\comp(1)} \exp\left[ 2 p^2 \frac{H^2}{f^2} G_\comp(\xi)\right]\,,
\end{equation}
where the constraint $p+q=0$ is ensured by the gauge symmetry of the compact scalar and $G_\comp(1)$ is the propagator at coincident points that arises from self-contractions (note this is not the normal ordered vertex operator, yet). Note that $G_\comp(\xi)$ is the (dimensionless) massless two-point function which in four dimensions is 
\begin{align}
        G_\comp(\xi) &= \frac{1}{8\pi^2} \left[\frac{1}{1-\xi}+\log\left(\frac{2}{1-\xi}\right)\right]\notag  \\[4pt]
        &= \frac{1}{8\pi^2} \left[\frac{2 \eta \eta'}{x^2-(\eta-\eta')^2}+\log\left(\frac{4 \eta \eta'}{x^2 - (\eta-\eta')^2}\right)\right]\,,
        \label{eq:Gcompxi}
\end{align}
where in the second line we have specialized to the dS flat slicing, and we are omitting the $i \epsilon$ prescription since it is not essential here. This is, strictly speaking, the Euclidean zero-mode subtracted propagator.
The necessity of the subtraction can be seen by integrating the standard equation of motion $\nabla^2 G_\comp(x,y)=\delta^{d+1}(x-y)$ on the Euclidean sphere. Since $\nabla^2 G_\comp(x,y)$ produces a pure boundary term which amounts to zero, this equation cannot be inverted \cite{Tolley:2001gg}. The origin of both terms in the position space expression \eqref{eq:Gcompxi} can also be derived from $k$-space as well. The equal time power spectrum of a massless field in dS is $P_\comp(\eta,k)= \frac{1}{2 k^3}+\frac{\eta^2}{2k}$.  Upon taking the Fourier transform of this expression, we can see from dimensional analysis that the $k^{-3}$ term is log divergent and therefore produces the $\log(\eta/x)$ term, whereas the $k^{-1}$ term (which is simply the conformal scalar power spectrum) produces the $\eta^2/x^2$ term.

It is also convenient to define
\begin{equation}
    \beta \equiv \frac{\Gamma(\tfrac{d}{2})}{4\pi^{\frac{d+2}{2}}} \frac{H^2}{f_\comp^2} \xrightarrow{d \to 3} \frac{H^2}{8\pi^2 f_\comp^2}\, .
    \label{eq:betaDef}
\end{equation}
 Normal ordering constitutes a removal of this divergent multiplicative constant, which is convenient to do in dimensional regularization. Hereafter, we will specialize to four dimensions. In physical correlators, the normal ordering constant can always be absorbed by a coupling constant, so we will simply work with the vertex operators $\mathcal{V}_p$. The 
 two-point function is then 
\begin{align}
      \big\langle \mathcal{V}^{\vphantom\dagger}_p(x) \mathcal{V}_p^{\dagger}(y)\big\rangle &=\left(\frac{1-\xi}{2}\right)^{-p^2 \beta} \exp\left(\frac{p^2 \beta}{1-\xi}\right) \notag\\[4pt]
        &= 2^{p^2 \beta}\sum_{n=0}^{\infty} \frac{(p^2 \beta)^n}{n!}\frac{1}{(1-\xi)^{p^2 \beta + n}} \,.
\label{eq:vertex_prop_bulk}
\end{align}
This vertex propagator in dS is unique in that is suffers from an essential singularity at short distances, unlike any other known operator which at worst only suffers from a polynomial singularity. A consequence of this singularity structure is that this correlator is actually \textit{regular} if the two points collide following a time-like path in spacetime, which corresponds to the limit $\xi \to 1^{+}$.

In the infrared, on the other hand, the vertex propagator breaks down into a sum of power laws, as is evident from (\ref{eq:vertex_prop_bulk}). Specifically, the equal time two-point function on the flat-slicing is 
\begin{equation}
    \big\langle \mathcal{V}^{\vphantom\dagger}_p(\eta,\vec{x}\s) \mathcal{V}_p^{\dagger}(\eta,\vec{0}\s)\big\rangle = 2^{p^2 \beta}\sum_{n=0}^{\infty} \frac{(2 p^2 \beta)^n}{n!} \left(\frac{\eta^2}{x^2}\right)^{p^2 \beta + n}\,.
    \label{eq:EqualTime2PointFlatSpaceSlicing}
\end{equation}
The expression in flat slicing will offer a useful match to the SdSET description of the vertex operators in Section~\ref{sec:sdset}.

\subsection{Lorentzian Inversion}

Let us now discuss the theory with an interaction between the principle series field $\phi$ and the compact scalar $\theta$, with the associated vertex operator $\mathcal{V}_p$.
For simplicity, we will only consider operators with $p=1$ and we will use the shorthand $\mathcal{V} \equiv \mathcal{V}_1$. The action is 
\begin{subequations}
\label{eq:phiVaction}
\begin{align}
 S &= \int 
    \d^{d+1}
    x\,\sqrt{-g} \big[\tfrac{1}{2}\partial_\mu \phi \partial^\mu \phi - \tfrac{1}{2}m_\phi^2 \phi^2 + \tfrac{1}{2}f_\comp^2 \partial_\mu \comp \partial^\mu  \comp \big] + S_{\rm int} \,,
\label{eq:int}
\end{align}
with
\begin{align}
    S_{\rm int} &= \int \d^{d+1}
    x\, \sqrt{-g}\,g_{\comp}\s \phi \cos(\comp) 
    \,\,\to\,\, \int \d^{d+1}x\, \sqrt{-g}\,\tfrac{1}{2}g_{\comp}\s \phi\left[\mathcal{V} + \mathcal{V}^{\dagger}\right]\,,
\end{align}
\end{subequations}
where it is understood that the coupling $g_\comp$ has absorbed a divergent factor due to a normal ordering prescription. From the discussion in Section~\ref{sec:setup}, the anomalous dimension of $\phi$ is now a straightforward application of \eqref{eq:anomalous_dim_from_spec} and is given by $\gamma_\phi= \frac{g_\theta^2}{4 \nu_\phi^2}\rho_\mathcal{V}(\Delta_\phi)$. We now need to determine the spectral density of the vertex operator.

\begin{figure}[t!]
    \centering
    \includegraphics[width=0.5\linewidth]{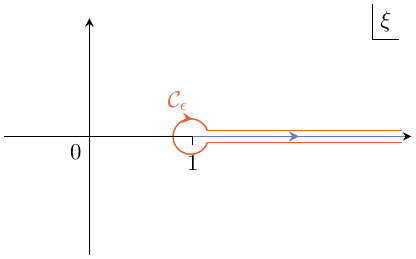}
    \caption{We illustrate two possibilities for the integration contour on the $\xi$-plane over which we could evaluate the Lorentzian inversion integral. The blue line indicates the choice where we evaluate the discontinuity first and then send $d \to 3$.  The red contour is the one we obtain if we set $d=3$ first. The difference between the two is essentially a residual integral over short-distances, which is denoted by the small circular contour $\mathcal{C}_\epsilon$.}
    \label{fig:lorentzian_contour}
\end{figure}

As discussed in \cref{sec:spectral_general} the spectral density can be obtained by inverting the KL representation. Let us first attempt to do so using Lorentzian data. The inversion integral reads 
\begin{equation}
    \rho_\mathcal{V}(\Delta) = \mathcal{N}_{\rm L}(\Delta) \int_1^{\infty} \ud \xi\, \tFo{1-\Delta}{1-\bar{\Delta}}{\tfrac{3-d}{2}}{\frac{1-\xi}{2}} {\rm disc}\,G_{\mathcal{V}}(\xi)\,,
\end{equation}
where $G_\mathcal{V}(\xi)\equiv \langle \mathcal{V}(x)\mathcal{V}^\dagger(y) \rangle$, and we have kept the number of dimensions arbitrary, with the goal of evaluating this expression for $d=3$. Note that we have encountered an order of limits ambiguity in this expression. We could either choose to place the contour onto the propagator's branch cut and then send $d \to 3$ (call this L) or vice versa (call this $\text{L}+\mathcal{C}_\epsilon$). The two operations do not commute.  In fact, for the latter option we cannot fully deform the integration contour due to an obstruction from the short-scale $\xi=1$ singularity of the propagator, leaving a small circular integral ($\mathcal{C}_\epsilon$) to be performed in addition. We show these two possibilities in \cref{fig:lorentzian_contour}. We will argue in what follows that $\text{L}+\mathcal{C}_\epsilon$ gives the correct answer.

To address this ambiguity, let us choose the option where we send $d \to 3$ after evaluating the discontinuity (the blue line in \cref{fig:lorentzian_contour}).  Using the form of the Lorentzian inversion formula given in \eqref{eq:lorentz_inv}, this calculation yields
\begin{equation}\label{eq:lor_naive}
    \rho_\mathcal{V}(\Delta) = \int_{1}^{\infty} \ud \xi\, \mathcal{K}_{\Delta}(\xi)\,{\rm disc}\,G_\mathcal{V}(\xi) \,,
\end{equation}
where we define the integration kernel
\begin{equation}
    \mathcal{K}_\Delta(\xi) \equiv 2 \pi ^2 i(2 \Delta -3) \cot (\pi  \Delta ) (\xi -1) \tFo{\Delta-1}{2-\Delta}{2}{\frac{1-\xi}{2}} \,,
\end{equation}
and the discontinuity of the propagator can be evaluated to yield
\begin{equation}
    {\rm disc}\,G_\mathcal{V}(\xi) = 2i \sin (\pi  \beta ) \left(\frac{\xi -1}{2}\right)^{-\beta } \exp\left(-\frac{\beta }{\xi -1}\right)\, ,
\end{equation}
where $\beta$ is the parameter defined in \eqref{eq:betaDef} and we evaluated this discontinuity using \eqref{eq:vertex_prop_bulk}.
The essential singularity of the vertex propagator ensures that the integral converges in the UV ($\xi \to 1$) for any choice of $\beta>0$.\footnote{To ensure convergence in the IR ($\xi \to \infty$), we require the absence of complementary series states, or $\beta > \frac{3}{2}$. In order to numerically evaluate the inversion integral for $\beta \leq \frac{3}{2}$ one could explicitly subtract off the complementary series contribution from the two-point correlation function inside the integrand. In such situations, if it is possible to evaluate the integral analytically, we could also continue our answer from $\beta > \frac{3}{2}$. However, this subtlety targets the correct inclusion of complementary series states. Since we are investigating the correct inclusion of principal series states, this is not relevant for our discussion. We will therefore assume $\beta> \frac{3}{2}$ while displaying numerical results.} 

If we take the second contour prescription, for which we set $d=3$ at the outset (the orange line in \cref{fig:lorentzian_contour}), we are no longer free to ignore the contribution from the deep UV, i.e.~around $\xi=1$. We are forced to include an extra contribution 
\begin{equation}\label{eq:lor_circ_corr}
    \delta\rho_\mathcal{V}(\Delta) = \int_{\mathcal{C}_\epsilon} \ud \xi\,\mathcal{K}_\Delta(\xi)\,G_\mathcal{V}(\xi)\,,
\end{equation}
where the contour $\mathcal{C}_\epsilon$ is parametrized as $\xi=1+\epsilon\s e^{i t}$ where $t \in (\epsilon, 2\pi-\epsilon)$. Strictly, this integral must be evaluated as we send $\epsilon \to 0^+$, and crucially this contribution is non-zero in this limit. In practice, we find that it is sufficient to keep $\epsilon \simeq 0.1$ for convergence and smaller values can contaminate the numerical integral due to severe oscillations of the integrand. We denote the inclusion of this extra UV contribution by ${\rm L}+ \mathcal{C}_\epsilon$.
\begin{figure}
    \centering
    \includegraphics[width=0.7\linewidth]{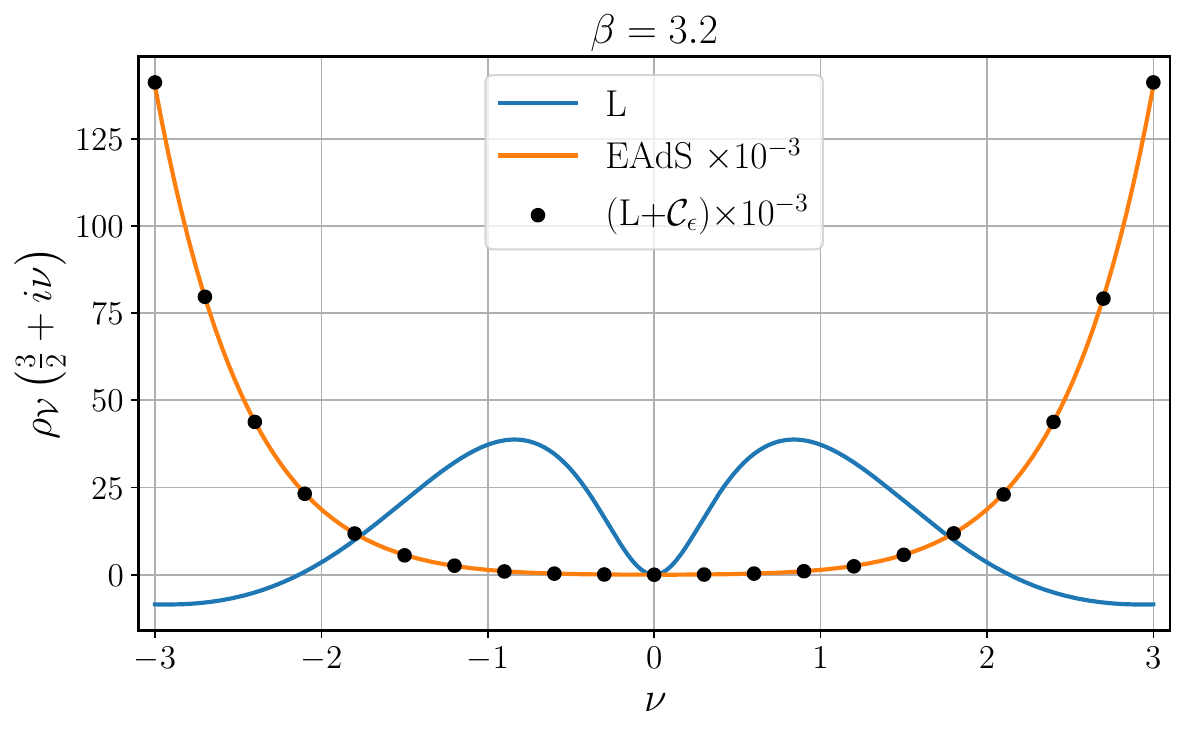}
    \caption{We show the result of the two choices for evaluating the Lorentzian inversion formula for a vertex operator, ${\rm L}$ and ${\rm L}+\mathcal{C}_\epsilon$. We pick $\epsilon=0.1$ in order to evaluate (\ref{eq:lor_circ_corr}), which we find is sufficient for convergence. Note that we have scaled ${\rm L}+\mathcal{C}_\epsilon$ by a factor of $10^{-3}$ in order to enable a visual comparison. We observe that the result of (\ref{eq:lor_naive}), denoted in blue, turns around and takes on negative values for some choices of $\nu$, apparently violating unitarity, whereas the inclusion of the short-distance information restores positivity. We also show the analytical result obtained using the EAdS inversion formula in solid orange which agrees with ${\rm L}+\mathcal{C}_\epsilon$.}
    \label{fig:compact_spectral_density}
\end{figure}

We show the spectral density obtained using both options in \cref{fig:compact_spectral_density}. Specifically, we see that only evaluating (\ref{eq:lor_naive}) can result in a negative spectral density, which is a clear violation of unitarity. 
Yet another point of difference is that ${\rm L}$ decays as $\nu \to \infty$ whereas ${\rm L}+\mathcal{C}_\epsilon$ grows. At large $\nu$, we are probing the UV behavior of the operator and we therefore expect to recover the flat-space spectral density. In flat-space, the spectral density $\rho_{\mathcal{V}}(\mu)$ is calculable as a sum over massless loops, from which one can extract the large $\mu$ asymptotic form \cite{Chakraborty:2025myb}:
\begin{equation}
    \log\rho_\mathcal{V}(\mu) \sim \left(\frac{\mu}{f_\comp}\right)^{\frac{2}{3}}\,,
\end{equation}
where the faster-than-polynomial growth is due to the fact that the vertex propagator exhibits an essential short-distance singularity in configuration space, as mentioned above. 
We conclude that the apparent violation of unitarity is due to a missing short-distance contribution, which is computed by $\mathcal{C}_\epsilon$.

Lastly, we can also turn to the EAdS inversion integral \eqref{eq:eads_inversion}. Note that EAdS does not see coincident points and is therefore free from any ambiguities we have encountered on the Lorentzian side. Moreover, \eqref{eq:eads_inversion} for a vertex operator can be integrated analytically in four dimensions \cite{Chakraborty:2025myb}. This is primarily possible due to the observation that for pairs of points accessible to EAdS, i.e.~$\xi \in (-\infty, -1)$, the vertex propagator can be expressed as an infinite tower of CFT states in the de Sitter bulk, $\Phi_{\delta}$, with scaling dimension $\beta+n$
\begin{equation}
    G_\mathcal{V}(\xi) = 2^{\beta} \sum_{n=0}^{\infty} G_{\beta+n}(\xi)\,.
\end{equation}
The resulting spectral density is 
\begin{align}
        \rho_\mathcal{V}(\Delta) = \,&\frac{16\pi \Gamma(1-\beta) \sin(\pi \beta)}{\Gamma(\beta-1)}(3-2\Delta)\cos(\pi \Delta)\notag \\
        &\times \Gamma(\beta-\Delta)\Gamma(\Delta-3+\beta)\,\tFt{\beta-\Delta}{\Delta-3+\beta}{\beta-1}{\beta}{\frac{\beta}{2}} \,.
\end{align}
We show both the result of this analytical expression and ${\rm L}+\mathcal{C}_\epsilon$ in \cref{fig:compact_spectral_density}, finding excellent agreement.  
Taken together, this justifies the statement claimed above that $\text{L}+\mathcal{C}_\epsilon$ gives the correct answer.  Not only does the calculation taking L alone violate positivity, but additionally it does not agree with the computation performed using EAdS. 

Therefore we conclude that the anomalous dimension $\gamma_\phi$ must unambiguously be positive, and an apparent negative sign originates from improperly accounting for UV data. In the next section, we will show how to reproduce this anomalous dimension using the long wavelength EFT.

\section{SdSET for Compact Scalars}\label{sec:sdset}

In this section, we will reconsider the impact of our periodic massless compact scalar field, $\comp$, now within the context of SdSET. We will again couple it to a principal series field, $\phi$, via the vertex operator $\mathcal{V}$, i.e. the action given in \eqref{eq:phiVaction}. We will show how the anomalous dimension for $\phi$ is captured using the long distance description.

\subsection{Vertex Operators in SdSET}
Following \cref{sec:SdSET}, we decompose the massless periodic UV field $\theta$ into two SdSET modes: 
\beq
\theta =H \big( \theta_+ + (aH)^{-3} \theta_-  \big) \, .
\eeq
Since $\theta$ massless scalar, the dimension of $\theta_+$ is $\Delta^{(\theta)}_+=0$ and therefore all polynomial operators of the form $\theta_+^n$ have the same power counting. This behavior is reminiscent of 2d CFTs and leads to a natural analogy with vertex operators in that context. While this may sound identical to the discussion in \cref{sec:inversion}, we emphasize that here our interest is in understanding how to think about vertex operators constructed using the SdSET mode $\theta_+$, not the UV field $\theta$.

Because the composite operators are naively degenerate, defining the correct basis of scaling operators requires understanding the mixing under RG evolution between $\theta_+^m$ and $\theta_+^n$ for any $n$ and $m$. As shown in \cite{Burgess:2015ajz,Gorbenko:2019rza,Baumgart:2019clc,Mirbabayi:2019qtx,Cohen:2021fzf,Mirbabayi:2020vyt,Burgess:2009bs,Prokopec:2017vxx,Cespedes:2023aal}, this mixing is equivalent to the formalism of stochastic inflation. Specifically, under time evolution the operators will mix according to  
\beq\label{eq:operator_ev}
\frac{\partial}{\partial \t } \langle \theta_+^N \rangle  =\sum_n g_n\binom{N}{n}\langle \theta_+^{N-n} \rangle\, ,
\eeq
where we imposed that this mixing equation for $\theta_+$ obeys a continuous shift symmetry~\cite{Cohen:2021jbo}. It is important that here we are not making any assumptions that the theory is free, as (shift-symmetric) integrations give rise to $g_{n>2} \neq 0$.

We can assume these correlators of $\tp$ are generated by a probability distribution, $P(\tp)$, such that 
\beq
\langle  F(\tp) \rangle = \int \d \tp F(\tp) P(\tp) \ .
\eeq
We can find a master equation for $P(\tp)$ using Equation~(\ref{eq:operator_ev}) and integrating by parts. This gives rise to a Fokker-Planck equation, 
\beq\label{eq:FP_gen}
\frac{\partial}{\partial \t } P(\theta_+) =\sum_n g _n \frac{(-1)^n}{n!} \frac{\partial^n}{\partial \theta_+^n} P(\theta_+)  \equiv \hat L_{\vp} P(\tp) \, .
\eeq
The fact that $\theta_+$ obeys a continuous shift symmetry forbids any explicit $\theta_+$ dependence in the Fokker-Planck equation beyond the derivatives~\cite{Cohen:2021jbo}. Given this master equation, we can now see that the operators 
\begin{align}
{\cal V}_{+,p} = e^{i p \theta_+}
\end{align}
are also scaling operators 
\beq
\frac{\partial}{\partial \t } \langle e^{i p \theta_+} \rangle  = \sum_n g_n (i p)^n \langle e^{i p \theta_+} \rangle  \equiv - \Delta_{p,+}\langle e^{i p \theta_+} \rangle\, .
\eeq
Since $\tp$ commutes with itself, we do not need to worry about normal ordering for ${\cal V}_{+,p}$. We notice that $\V_{p,+}$ is a scaling operator as a result of the shift symmetry for $\theta_+$ and does not dependent on the specific interactions that given rise to $g_n\neq 0$. In the free theory,
\beq
g_{n\neq 2}=0 \qquad  \text{and} \qquad g_2 = \frac{1}{8\pi^2} \frac{H^2}{f_\theta^2 } \equiv \beta\,,
\eeq
so that $\Delta_{p,+} = p^2 \beta$, in agreement with \cref{sec:inversion}. It should also be clear that the generalization to interacting theories will leave the results largely the same and only change the specific value of $\Delta_{p,+}$.

The two point correlation functions of these operators at separated points can also be determined within the stochastic formalism~\cite{Starobinsky:1994bd}. The most straightforward calculation is the two point function of a time-like separated correlator. We can express this as an integral over the two-point probability distribution function (pdf)
\beq
 \langle \V_{p,+}(\t) \V_{p',+}(\t') \rangle =\int \d \tp \d\tp' e^{i 2\pi p \tp + 2\pi p' \tp'} \rho_2\big[\tp( \t )\mid \tp'( \t') \big] \, .
\eeq
We have dropped the dependence on the comoving position $\x$ because we have assumed both operators are at the same point. The joint probability distribution is typically written in terms of transition amplitudes as
\begin{align}
\rho_2\left[\tp( \t )| \tp'( \t') \right] &=  \Pi\big[\tp( \t ) \mid  \tp'( \t')\big] \rho_1\big[\tp'( \t')\big] \theta(\t-\t' ) \notag \\[4pt]
&\hspace{12pt} +\Pi\big[\tp'( \t' ) \mid  \tp( \t)\big] \rho_1\big[ \tp( \t) \big] \theta(\t'-\t ) \, ,
\end{align}
where $\Pi[A|B]$ is the conditional probability of $A$ given $B$, and $\rho_1$ are the one-point distribution functions.
Since the evolution of the probability distribution is Markovian, the transition amplitudes must also obey the generalized Fokker-Planck equation in Equation~(\ref{eq:FP_gen}), 
\beq
\frac{\partial}{\partial \t}  \Pi\big[\tp(\t ) \mid  \tp'(\t')\big] = \hat L_{\vp} \Pi\big[\tp( \t ) \mid  \tp'(\t')\big] \,,
\eeq
and similarly for the derivative with respect to $\t'$. Because $ \V_{p,+}(\t)$ is an eigenvalue of $\hat L^\dagger$ (the operator after integration by parts). Recall that this conclusion applies even to interacting theories, as long as they obey the shift symmetry in $\tp$. Now we can conclude that 
\beq
\langle \V_{p,+}(\t) \V_{p',+}(\t')  \rangle =A(\t') \exp( -\Delta_{p,+} \t) \theta(\t -\t') + B(t_2) \exp( -\Delta_{p',+} \t') \theta\big(\t' -\t\big) \,.
\eeq
To determine the unknown functions, we require that when $\t = \t'$, this two-point function should reproduce the equilibrium one-point pdf $\rho_1(\tp) = {\rm constant}$, which is possible because the scalar is compact. Integrating over $\vp$ in the equilibrium state gives $\delta_{j,-j'}$ and therefore
\beq
\langle \V_{p,+}(\t) \V_{p',+}(\t')  \rangle =\delta_{p,-p'} \exp( -\Delta_{p',+}| \t-\t'|)  \ .
\eeq
We see that stochastic inflation gives us the expected two-point statistics for an operator $\V_{p,+}$ that has dimension $\Delta_{p,+}$. One can apply these same techniques at spacelike separation to show the stochastic formalism produces the dS-invariant results consistent with QFT expectations~\cite{Starobinsky:1994bd,Finelli:2008zg,Vennin:2015hra,Gorbenko:2019rza}. This shows that stochastic inflation reproduces the non-trivial leading order long distance behavior but does not capture the subleading terms in \eqref{eq:vertex_prop_bulk}.

\subsection{Subleading Behavior of Vertex Operators}

Unfortunately, the information we need to find the anomalous dimension of $\phi$ is not limited to the $\Delta_+ =0$ sector of $\theta$ that is described by stochastic inflation alone. The problem is that the stochastic formalism only captures the leading time-dependence of $\theta_+$ and is entirely independent of $\theta_-$. However, we know that both $\theta_+$ and $\theta_-$ along with derivative corrections to both fields are required to capture the long wavelength dynamics of $\theta$. 
Therefore, in order to match the direct calculation, it is important to realize that $\V_p \neq \V_{p,+}$. Specifically, the UV field decomposes into a infinite list of operators,
\beq\label{eq:SdSET_derivatives}
\theta = \theta_+ -\frac{1}{2} (aH)^{-2} \partial^2 \theta_+ +(aH)^{-3} \theta_- - \frac{1}{8} (aH)^{-4} \partial^4 \theta_+ +\cdots \ .
\eeq
Here we are making a departure from the SdSET ansatz by introducing the gradients of $\theta_+$ explicitly\footnote{This procedure of isolating the derivatives is straightforward at a single time $\t$. The time-evolution of the operators to other times is then understood via RG flow.} and, in principle, gradients of $\theta_-$, so that $\langle \theta_+ (\k\s)\theta_+(-\k\s)\rangle' \propto k^{-d}$ with no higher powers of $k$.  Now the vertex operator of the UV field becomes
\begin{align}
\V = \exp(ip \theta) &= \exp\!\Big(ip \big( \theta_+ + (aH)^{-2}\partial^2 \theta_+  +(aH)^{-3} \theta_- + \cdots\big) \Big) 
\notag\\[3pt]
&= \V_{p,+} \left(1+(aH)^{-2}\partial^2 \theta_+ +i p (aH)^{-3} \theta_- +\cdots \right)\,.
\end{align}
Technically speaking, this definition requires a specific ordering of $\theta_+$ and $\theta_-$, which is ambiguous: if we define the behavior at coincident points using dimensional regularization,\footnote{Recall that the cutoff of the EFT is at $k=aH$, so that $[\theta_+(\x) ,\theta_-(\x)] = i \int \d^3 k  = (aH)^3$. This cancels with the $(aH)^{-3}$ multiplying $\theta_-$ to give an time-independent correction that can be absorbed into the definition of $\V$} then $[\theta_+(\x) ,\theta_-(\x)] = 0$ and there is no such ambiguity. 

We are still dealing with the free theory, so at this point the only anomalous dimensions are associated with $\theta_+$, via $\V_{p,+}$ and therefore we get an infinite series of terms that have effective dimensions $\Delta_{p,+} = p^2 \beta$, $\Delta_{p,+}+2$, $\Delta_{p,+}+3$, etc. It is also important that operators like $\partial^2 \theta_+$ are ``descendants" in the sense of the conformal Ward identities. Specifically, we know that in free theory,
\beq
\langle \theta(\k,\t) \theta(\kp,\t)\rangle' = k^{-3}(1 + (aH)^{-2} k^{2}) \ . 
\eeq
The second term arises from $\langle \partial^2 \theta_+(\k\s) \theta_+(\kp) \rangle = k^{-1} \neq 0$. The resulting two point function is the same as if we introduced an operator of dimension 1, but it is important that there is not a primary operator of dimension 1 in SdSET, only these descendants. One can check that the contribution from $\partial^4 \theta_+$ cancels with the $\langle \partial^2 \theta_+ \partial^2\theta_+ \rangle $ term. This pattern will continue as we add higher derivative terms to \eqref{eq:SdSET_derivatives}. In this sense, $\theta$ must contain an infinite series of operators (descendants) even though the power spectrum only contains two terms.

We see that in SdSET, the UV vertex operator decomposes into an infinite sum of vertex operators dressing higher dimension local operators made from $\tp$, $\tm$, and derivatives. All that remains is to perform a matching calculation to determine that relationship between the UV coupling and the EFT couplings (as usual, one UV coupling becomes infinitely many EFT couplings). For the purposes of a one-loop anomalous dimension calculation, it most straightforward to match the equal-time two point statistics of $\V_p$ at separated points. Since the theory is Gaussian, we know
\begin{align}
\hspace{-8pt}\langle \V_p(\x\s) \V_{-p}(0) \rangle &= \langle \exp(2\pi i p (\theta(\x\s) -\theta(0)) \rangle = \exp\!\big(-2 \pi^2 p^2 \langle (\theta(\x\s) -\theta(0))^2 \rangle \big) \\[4pt]
&= \langle \V_{p,+}(\x\s) \V_{-p,+}(0) \rangle \exp\!\big(-2 \pi^2 p^2(\langle \partial^2 \tp(\x\s) \tp(0) \rangle+ \langle  \tp(\x\s) \partial^2 \tp(0) \rangle) \big) \, . \notag
\end{align}
Given this relation, we can identify all the operators relevant to the correlations at space-like separated points and then use dS invariance to determine the full correlation functions of those operators.

However, the key issue, as discussed in \cref{sec:SdSET}, is what happens at coincident points in space, where $\x =0$, but not necessarily equal times. Specifically, many of these correlators contain contact terms in the UV description, even though they are naively forbidden by dS invariance. For example, an operator ${\cal O}_{p,n} = \V_{p,+} (\partial^2 \tp)^n$ has dimension $\Delta_{p,n} = \Delta_{+,p} + 2n$. The two point function of ${\cal O}_{p,n}$ cannot contain a contact term, but we can introduce a shadow operator ${\bar {\cal O}}_{-p,n}$ with dimension $\bar \Delta_{p,n} = d -\Delta_{p,n} $ so that 
\beq
\langle {\cal O}_{p,n}(\x,\t) {\bar {\cal O}}_{-p,n}(0,\t') \rangle =C(\Delta_{p,n}) (aH(\t))^{-\Delta_{p,n}} (aH(\t'))^{-\bar \Delta_{p,n}} \delta(\x\s) \, .
\label{eq:shadownContactTerm}
\eeq
We determine can determine all coefficients $C(\Delta_{p,n}) $ by matching to the UV description.  We do so by matching the position space power-law dependence.

\subsection{Matching SdSET Contact Terms to a Bulk CFT}
A pragmatic approach to keeping all of the SdSET operators that appear in the UV vertex operator is to expand in powers (scaling operators). Specifically, we can write the two-point function of the vertex operator, \eqref{eq:EqualTime2PointFlatSpaceSlicing}, as an infinite sum of bulk ``CFT operators" $\Phi_\delta$ with dimension $\delta = \beta+n$. At equal times and super-horizon distances, we can trivially define ``operators" to represent the powers in the series expansion, so that
\begin{equation}
\langle \mathcal{V}(\x_1)\mathcal{V}^\dagger(\x_2)\rangle=2^\beta \sum_{n=0}^{\infty} \frac{\beta^n}{n!} \langle \Phi_{\beta+n}(\x_1)\Phi_{\beta+n}(\x_2)\rangle\, ,
\end{equation}
where
\beq
\langle \Phi_{\beta+n}(\x_1,\eta)\Phi_{\beta+n}(\x_2,\eta)\rangle \equiv \frac{2^{\beta+n}(-\eta)^{2(\beta+n)}}{|\x_1-\x_2|^{2(\beta+n)}} \ .
\eeq
Of course, to perform calculations, we generally need to know the correlators at unequal times. However, since we know the answer is dS invariant, we can write our result in terms of the two-point correlation function for a bulk CFT with dimension $\delta$, 
\begin{equation}
 \langle \Phi_{\delta}(\x_1, \eta_1)\Phi_{\delta}(\x_2,\eta_2)\rangle   \equiv G_\delta(\x_1-\x_2,\eta_1,\eta_2)=\frac{2^\delta\left(-\eta_1 \right)^\delta\left(-\eta_2\right)^\delta}{\left[(\x_1-\x_2)^2-\left(\eta_1-\eta_2\right)^2\right]^\delta}\,.
\end{equation}
Here $\Phi_\delta$ does not need to live in a known interacting CFT, rather it is just statistical variable with specified power-law behavior. The normalization, $C(\Delta)$, is determined by matching.

At this point, the strategy will be develop some intuition by studying how a single scaling operator generates an anomalous dimension, and then summing over all the contributions according to the known formula for the series expansion. For an interacting term of the form \eqref{eq:int}, the anomalous dimension is
\begin{equation}\label{eq:vertex_from_CFT}
\gamma_{\phi} = 2^\beta\sum_n \frac{\beta^n}{n!}\gamma_{\delta = \beta +n}\,,
\end{equation}
where $\gamma_\delta$ denotes the anomalous dimension of $\phi$ when we have interactions of the form
\begin{equation}
  S_{\rm int} = \int \ud^{d+1}x\, \sqrt{-g}\,g_{\comp}\s \phi\s \Phi_\delta\,.
\end{equation}
This representation will prove to be particularly useful for matching. 

Loop calculations will generally be easier using the momentum space propagator, which takes the form
\begin{equation}
  G_\delta(\k,\eta_1,\eta_2)   =  \frac{(2\pi)^{d/2}}{\Gamma(\delta)}(-\eta_1)^\delta \left(-\eta_2\right)^\delta\left(\frac{k}{i(\eta_1-\eta_2)}\right) ^{\delta-d/2} K_{d/2-\delta}\big(i k (\eta_1-\eta_2)\big)\,,
\end{equation}
where $K_\alpha(z)$ is the modified Bessel function of the second kind of order $\alpha$. The modified Bessel function $K_{d/2-\delta}$ admits a series representation that allows us to read off the contact terms immediately
\begin{equation}\label{eq:contact}
 G_\delta(\k,\eta_1,\eta_2) \stackrel{k^0}{=} 
e^{-i\pi(d/2-\delta)}\frac{\pi^{d / 2} 2^{\delta} \Gamma\left(\delta-\frac{d}{2}\right)}{\Gamma(\delta)}\left(-\eta_1\right)^\delta \left(-\eta_2\right)^\delta\left(|\eta_2-\eta_1|^2\right)^{d/2-\delta}.
\end{equation}
We note that the equal time contact term diverges for $\delta > d/2$, which leads to a divergent time integral. A similar problem has also been observed near the boundary of AdS. In both cases, the problem can be 
avoided by a choice of
analytic continuation\cite{Bzowski:2015pba}. We review the details of how to perform analytic continuation in the UV theory for a bulk CFT with $\delta > \frac{d+1}{2}$ in \cref{app:In-in}. 

Since we are interested in the unequal time contact terms for SdSET, we can rewrite \eqref{eq:contact} as 
\begin{equation}
 G_\delta(\k,\eta_1,\eta_2) \stackrel{k^0}{=} \theta(\eta_2-\eta_1)
\sum_n i \frac{2^{1+d-\delta} e^{i \pi \delta} \pi^{\frac{1+d}{2}} \Gamma(-d+n+2 \delta)}{\Gamma(n+1) \Gamma(-\frac{d-1}{2}+\delta) \Gamma(\delta)}\left(-\eta_1\right)^{d-\delta-n}\left(-\eta_2\right)^{\delta+n} \ .
\end{equation}
This result is identical to the contact term obtained using the spectral density for a bulk CFT\cite{Hogervorst:2021uvp}.
Following \eqref{eq:shadownContactTerm}, we include operators $\mathcal{O}_{\delta+n}$ and $\bar\O_{\delta+n}$ with the following commutation relation in SdSET
\beq
\langle[ \bar\O_{\delta+n}(\t), \O_{\delta+n}(\t')]\rangle = -\theta(\t-\t')\frac{\pi ^{\frac{d+1}{2}} 2^{d-\delta +2} \sin \left(\frac{1}{2} \pi  (d-2 \delta
   )\right) \Gamma (-d+n+2 \delta )}{\Gamma(n+1)\Gamma(-\frac{d-1}{2}+\delta) \Gamma(\delta)} \ .
\eeq
The inclusion of $\O_{\delta+n}$ and $\bar\O_{\delta+n}$ means the anomalous dimension for $\phi$ in SdSET can be written as the following series sum 
\beq
\gamma_\phi + i \delta \nu_\phi =\frac{g^2}{4\nu_\phi}\sum_n\frac{ i \pi ^{\frac{d+1}{2}} 2^{d-\delta +3} \Gamma (-d+n+2 \delta )}{(n+\delta-\Delta_\phi)\Gamma(n+1) \Gamma (\delta
   ) \Gamma \left(-\frac{d}{2}+\delta +\frac{1}{2}\right)}\sin\left(\pi\left(\frac{d}{2}-\delta\right)\right) \ .
\eeq
In the limit $n\to \infty$, both the real and the imaginary part of the summand scale as $n^{2\delta-d-2}$,
which implies that for $\delta \geq \frac{d+1}{2}$, the series is necessarily divergent. We can obtain a finite sum via analytic continuation by rewriting the summand as (see \cref{app:In-in})
\begin{align}
&\gamma_\phi+i \delta \nu_\phi  =- \frac{i g^2}{4\nu_\phi}\frac{\pi ^{\frac{d+1}{2}} 2^{\tfrac{d}{2}-x +3}\Gamma(x+\tfrac{d}{2}-\Delta_\phi)\Gamma(2x)\sin(\pi x)}{\Gamma (x+\tfrac{d}{2}
   ) \Gamma\left(x +\frac{1}{2}\right)\Gamma\left(x+\tfrac{d+2}{2}-\Delta_\phi\right)}\sum_n \frac{1}{n!}\frac{(x+\tfrac{d}{2}-\Delta_\phi)_n(2x)_n}{(x+\tfrac{d+2}{2}-\Delta_\phi)_n}
   \notag\\[5pt]
   &=-\frac{i g^2}{4\nu_\phi}\frac{\pi ^{\frac{d+1}{2}} 2^{\tfrac{d}{2}-x +3}\Gamma(x+\tfrac{d}{2}-\Delta_\phi)\Gamma(2x)\sin(\pi x)}{\Gamma (x+\tfrac{d}{2}
   ) \Gamma\left(x +\frac{1}{2}\right)\Gamma\left(x+\tfrac{d+2}{2}-\Delta_\phi\right)}  \tFo{\delta-\Delta_\phi}{-d+2\delta}{\delta+1-\Delta_\phi}{1}\label{eq:CFT_sum}\, , 
\end{align}
where $x=\delta -\tfrac{d}{2}$ and $(a)_n$ is used to denote the Pochhammer symbol. To get from the first to the second line in \eqref{eq:CFT_sum}, we summed the series by applying the definition of the hypergeometric function. The hypergeometric function in \eqref{eq:CFT_sum} is given by
\beq\label{eq:hypergeometric_series_sum}
\tFo{\delta-\Delta_\phi}{-d+2\delta}{\delta+1-\Delta_\phi}{1} = \frac{\Gamma(\delta-\Delta_\phi+1)\Gamma(1+d-2\delta)}{\Gamma(1-\delta+d-\Delta_\phi)}, \quad 1+d-2\delta >0 \, ,
\eeq
which is divergent for $\delta > \frac{d+1}{2}$, but since the RHS of \eqref{eq:hypergeometric_series_sum} is finite for $\delta > \frac{d+1}{2}$, we can use it to analytically continue in $\delta$. This then yields the following anomalous dimension for $\phi$
\beq\label{eq:complex_sum_CFT}
\gamma_\phi + i\delta \nu_\phi = -i\frac{g^2}{\nu_\phi}\frac{\pi ^{\frac{d+1}{2}} 2^{d-\delta } \Gamma \left(\frac{d+1}{2}-\delta\right) \Gamma (\delta -\Delta_\phi )}{\Gamma (\delta )
   \Gamma (d-\delta -\Delta_\phi +1)}\, .
\eeq
This expression for the anomalous dimension has a non-vanishing imaginary part, which can be removed by including a mass counterterm following the on-shell renormalization scheme discussed in \cref{app:anom_dims_euclidean}. Taking the real part of \cref{eq:complex_sum_CFT} then leads to
\beq
\gamma_\phi = \frac{\pi ^{\frac{d+1}{2}} g^2 2^{d-\delta } \sin \left(\pi  \left(\Delta_\phi
   -\frac{d}{2}\right)\right) \Gamma (\delta -\Delta_\phi ) \Gamma (\delta -\bar\Delta_\phi
   )}{\left(\Delta_\phi -\frac{d}{2}\right) \Gamma (\delta ) \Gamma \left(-\frac{d}{2}+\delta
   +\frac{1}{2}\right)}\,,
\eeq
which reproduces the result using \eqref{eq:anomalous_dim_from_spec}.
\subsection{Anomalous Dimensions From Vertex Operators}

Given the matching between the vertex operators and a sum over CFT operators with specified contact terms, it is straightforward to calculate the anomalous dimension of $\phi$. For any given operator in the spectrum $\O_n$ with dimension $\Delta_n$ and shadow partner $\bar \O_n$, one finds a term of the form in \eqref{eq:SdSETlog}. Summing over $n$, or the spectrum of operators, yields the full logarithmic correction to the power spectrum of $\phi$:  
\beq 
\langle \vp(\k\s) \vp(\kp) \rangle' = k^{2 i \nu} \left(1+ g^2 \frac{\log (-k \eta)}{2 \nu}\sum_n \frac{C(\Delta_n)}{\Delta_n-\Delta_\phi}  \right)\, .
\eeq
The coefficient of this log is twice the anomalous dimension of one of these operators, so that
\beq \label{eq:anom_dim_from_sum2}
\gamma_\phi = \frac{g^2}{4\nu_\phi} {\rm Re} \sum_n \frac{C(\Delta_n)}{\Delta_n-\Delta_\phi} \ .
\eeq
Recall that in \cref{sec:connecting}, we showed that this formula is equivalent to the spectral density formula \eqref{eq:anomalous_dim_from_spec}. Specifically, the sum over operators $\O_n$ is just the residue formula for an contour integral over $\rho(\Delta)$.

In order to match the spectral density formula, we still require that the contour at infinity vanishes. This is the same as requiring that our sum in \eqref{eq:anom_dim_from_sum2} converges. This is rarely the case, in practice. Nevertheless, behavior of the sum at $\Delta_n \to \infty$ reflects the UV behavior of the theory and should removable by local counter-terms.

One way to obtain a finite result is to regulate the sum so that it converges. We can then express the two point function for the vertex operator as a sum over the propagators for bulk CFTs, as shown in \eqref{eq:vertex_from_CFT}. Then, term by term, we can regulate the contribution from each $\delta$, as we did in the previous section. The anomalous dimension for the exchange of a vertex operator is then the sum over these contributions:  
\beq
\gamma_\phi + i\delta \nu_\phi = \sum_{n} \frac{\pi ^{\frac{d+1}{2}} g^2 \beta ^n 2^{-\beta +d-n} \Gamma \left(\frac{d+1}{2}-
   (n+\beta )\right) \Gamma (n+\beta -\Delta_\phi )}{n! \left(\Delta_\phi -\frac{d}{2}\right)
   \Gamma (n+\beta ) \Gamma (\bar\Delta_\phi-n-\beta  +1)} \ .
\eeq
The sum over CFT operators is itself a perfectly convergent series as the summand scale as $2^{-n}/n!$ for large $n$. To sum the series, we can again rewrite the gamma functions in terms of Pochhammer symbol and apply the definition of the hypergeometric function. This then yields
\begin{align}
 \gamma_\phi &= \frac{\pi ^{\frac{d}{2}+\frac{1}{2}} g^2 2^{-\beta +d+1} \Gamma (\beta -\Delta_\phi )  \sin\!
   \big(\frac{1}{2} \pi  (d-2 \Delta_\phi )\big) \Gamma (\beta -\bar\Delta_\phi ) \,
   }{\Gamma (\beta ) (d-2 \Delta_\phi ) \Gamma
   \left(\frac{1}{2} (-d+2 \beta +1)\right)} 
   \tFt{\beta-\Delta_\phi}{\!\Delta-d+\beta}{\beta-\frac{1}{2}+\frac{d}{2}}{\beta}{\frac{\beta}{2}} \!.\notag\\
\end{align}
This also reproduces the result obtained using \eqref{eq:anomalous_dim_from_spec}.

Finally, with the presence of a real anomalous dimension $\gamma_\phi > 0$, the two point function of $\phi$ must also contain a (positive) contact term as discussed in \cref{sec:2.1}. Usually contact terms are ambiguous, unless they are fixed by symmetries. In this case, the situation is non-trivial because no such term would be allowed for operators with  ${\rm Re} \Delta \neq d/2$. However, the presence of the shadow operators presents an opportunity for such a term to arise. In fact, the contact term in the interacting theory can be derived using the spectral density for $\phi$ \cite{SalehiVaziri:2024joi}

\begin{equation}
    c = \int_{\frac{d}{2}-i\infty}^{\frac{d}{2}+i\infty} \frac{\ud \Delta}{2\pi i} \rho_{\phi}(\Delta) \frac{\coth(\pi \nu)}{2\nu}\ ,
\label{eq:contactterm}
\end{equation}
where the term $\frac{\coth(\pi\nu)}{2\nu}$ comes from the contact term for a free principal series field whose mass is fixed by $\nu$.

A priori, the convergence of this integral is not guaranteed, as the spectral density of a general operator may grow at large $\nu$. However, for interactions of the form \eqref{eq:general_interaction}, the spectral density for $\phi$ is calculable for specific examples of $\Phi$\cite{Loparco:2025azm}. We also discuss the details of this calculation in \cref{app:anom_dims_euclidean} and show that the spectral density $\rho_\phi(\nu) \to 0$ when $\nu \to \pm \infty$ in \cref{fig:example_corr_spectral_density}. Moreover, the decay of $\rho_\phi \to 0$ as $\nu \to \pm\infty$ can be attributed to the fact that the (renormalized) principal series field $\phi$ satisfies the canonical commutation relation. As discussed in \cite{Weinberg:1995mt}, the Weinberg sum rule tells us that 
\beq
    \int_{\frac{d}{2}-i\infty}^{\frac{d}{2}+i\infty} \frac{\d\Delta}{2\pi i}\rho(\Delta) = c_1 \, ,
\eeq
where $c_1$ is some finite positive constant that is fixed by the renormalization condition discussed in \cref{app:anom_dims_euclidean}.

Finally, we note that integrand of the contact term \eqref{eq:contactterm} has a term that diverges as $\coth(\pi \nu)/\nu \to \nu^{-2}$ as $\nu \to 0$, which could naively lead to a divergence. However, as we explained in Section~\ref{sec:connecting}, the $\Delta=\frac{d}{2}$ state is typically absent in the Hilbert space of scalar operators $\Phi$ and $\rho_\Phi(\nu)\to \nu^2$, which causes $\rho_\phi(\nu)\to \nu^2$ in the limit $\nu \to 0$ as well, eliminating this divergence.  This is consistent with the conclusion that the contact term $c$ must be a finite positive constant.

\section{Conclusions}\label{sec:conclusions}
Recent years have seen significant progress in understanding perturbative quantum field theory in de Sitter space. The dS isometeries provide a set of powerful constraints on the structure of cosmological correlators that can be used to organize calculations in ways that are familiar from AdS and flat space~\cite{Baumann:2022jpr}. In addition, EFT techniques~\cite{Green:2022ovz} have made the physics of the long wavelength modes manifest, enabling a variety of all-orders results. Concretely, by power counting, most interactions in dS are irrelevant, which ensures that the long wavelength theory is IR free (with the important exception of light scalar fields).

Despite this progress, the challenge remains to understand the uniquely cosmological features of dS space, inflation, and other time-dependent backgrounds. Although these theories can be organized in terms of familiar symmetries, soft theorems, etc, the theories do not enjoy the same non-perturbative organizing principles~\cite{Flauger:2022hie}, such as the $S$-matrix or the state operator correspondence. This makes interpreting and checking increasingly high order perturbative calculations difficult, especially when moving beyond static dS to theories with inflation and/or when including dynamical gravity. 

A step towards building a robust understanding of quantum cosmology is to develop connections between results that are derived using different approaches. In this paper, we developed several new connections between the EFT approach and the spectral representation of the correlators. In particular, we demonstrated that the results for the anomalous dimensions of principal series scalars agree upon properly accounting for renormalization. In the process, we established new results for the positivity of these anomalous dimensions. We explored the coupling of a principle series field to compact light scalars as a specific non-trivial example. We illustrated many subtleties that appeared for both the spectral density and EFT calculations that are required to find consistent results across both methods.

Understanding the precise relationship between perturbative calculations and non-perturbative principles is part of the larger goal of understand the full range of long distance physics that can arise in cosmological backgrounds. In pure dS, one might hope~\cite{Hogervorst:2021uvp} to develop a non-perturbative bootstrap approach similar to unitary CFTs~\cite{Poland:2022qrs,Achucarro:2010da}. Strict constraints from positivity are an essential ingredient to such a program, as the are ultimately the origin of bounds on dimensions of operators and OPE coefficients. 

A broader ambition is to understand the space of predictions for inflationary correlators. The current organization of searches for interactions is ultimately perturbative in nature~\cite{Babich:2004gb,Planck:2019kim}. A non-perturbative approach to classifying the space of inflationary signals is a concrete goal that is far beyond the reach of current techniques. Inflationary models with strongly broken dS isometeries are not known to obey many of the constraints implied by the spectral density in dS~\cite{Green:2023ids}. For related reasons, these models are often the most compelling observation targets~\cite{Chen:2006nt,Cheung:2007st,Flauger:2009ab}, making a broad classification relevant for the current and future data analysis program~\cite{Achucarro:2022qrl}. Through specific examples, it is known that the information about the inflationary epoch can leave an imprint on large multiplicity correlations. Yet, even in these examples, the long distance modes are generally amenable to a weakly coupled EFT treatment~\cite{Cohen:2021jbo,Green:2024fsz}, which suggests some aspects of the decomposition into scaling operators is likely to survive the change from dS to inflation.  By way of the compact scalar example, this paper makes it clear that we are only beginning to understand how to approach these critical outstanding issues systematically.  Continuing to develop and stress test various theoretical tools will pave the way to ensuring that all possible signatures of inflation are accounted in the analysis of current and future cosmological surveys.

\clearpage

\paragraph{Acknowledgments}
We are grateful to Kshitij Gupta, Austin Joyce, Enrico Pajer, Akhil Premkumar, Matt Reece, Eva Silverstein, John Stout, and Guanhao Sun.
TC is supported by the US~Department of Energy under grant~\mbox{DE-SC0011640}.
DG and YH are supported by the US~Department of Energy under grant~\mbox{DE-SC0009919}. PC is supported by the DOE Grant~\mbox{DE-SC0013607}.

\section*{Appendices}
\phantomsection
\addcontentsline{toc}{section}{Appendices}
\appendix


\section{In-In Time Integrals}\label{app:In-in}
The standard approach to cosmological correlators is via integrating the interaction picture Hamiltonian over cosmological time. In this appendix, we will discuss some aspects of these calculations that are relevant to the main text.

\subsection{Evaluating Time Integrals}

Before we get to physics, it is useful to first review how time-integrals can be regulated. A common time integral we encounter in the UV theory takes the form
\beq\label{eq:eta_integral}
\int_{-\infty}^{\eta_2} \mathrm{~d} \eta_1 \left(-\eta_1\right)^{-d-1+\Delta}(-\eta_1)^\delta (\eta_2-\eta_1)^{d-2\delta} \, .
\eeq
This integral can be evaluated via a change a variable $t =\frac{\eta_2}{\eta_1}$ and applying the following integral representation of the hypergeometric function  
\begin{equation}
{ }_2 F_1(a, b ; c ; x)=\frac{\Gamma(c)}{\Gamma(b) \Gamma(c-b)} \int_0^1 \d t\, t^{b-1}(1-t)^{c-b-1}(1-x t)^{-a} \, .
\end{equation}
For $\delta \geq \frac{d+1}{2}$, the integral in \eqref{eq:eta_integral} is divergent. When it converges, our integral is the hypergeometric function evaluated at $x = 1$, which is finite and given by
\begin{equation}\label{eq:2f1}
{ }_2 F_1(a, b ; c ; 1)=\frac{\Gamma(c) \Gamma(c-a-b)}{\Gamma(c-a) \Gamma(c-b)} \qquad {\rm when} \qquad \text{Re}(c)>\text{Re}(a+b) \, .
\end{equation}
Outside of this range, we can nevertheless define the integral for $\delta >\frac{d+1}{2}$ by making use of the above identity in \eqref{eq:2f1}, since the RHS is finite even if the condition $\text{Re}(c)>\text{Re}(a+b)$ is not met. As a consequence, the time integral in \eqref{eq:eta_integral} is
\beq
\int_{-\infty}^{\eta_2} \mathrm{~d} \eta_1 \left(-\eta_1\right)^{-d-1+\Delta}(-\eta_1)^\delta (\eta_2-\eta_1)^{d-2\delta} = (-\eta_2)^{\Delta-\delta}\frac{\Gamma(1+d-2\delta)\Gamma(\delta-\Delta)}{\Gamma(1+d-\delta-\Delta)}\,,
\eeq
for all $\delta$.
\subsection{In-In Time Integrals From the UV}
In flat slicing coordinates, we can calculate the corrections to the two point correlation function using in-in perturbation theory, following \cite{Weinberg:2005vy}, the $\mathcal{O}(g^{2})$ correction for an interaction of the form \eqref{eq:general_interaction} is 
\begin{align}
\langle\phi(\k,\eta)\phi(\kp,\eta)\rangle' &=
 -g^2 \int_{-\infty}^{\eta}\d\eta_2\s a(\eta_2)^{d+1}\big[\phi(\k,\eta_2),\phi(\kp,\eta)\big]    \notag\\
 &\hspace{15pt}\times\bigg( \int_{-\infty}^{\eta_2} \d\eta_1 a(\eta_1)^{d+1}\langle\phi(\kp,\eta_1)\phi(\k,\eta)\rangle \big\langle[\Phi(\k,\eta_1),\Phi (\kp,\eta_2)\big]\big\rangle\label{eq:anomalous_dim_in-in}\\[3pt]
&\hspace{40pt}+\int_{-\infty}^{\eta_2} \d \eta_1\s a(\eta_1)^{d+1} \big[\phi(\kp,\eta_1),\phi(\k,\eta)\big]\big\langle\Phi(\k,\eta_1)\Phi(\kp,\eta_2)\big\rangle \bigg) \, .\label{eq:positivity} 
\end{align}
where all terms inside the time integrals are evaluated at $g=0$. Both \eqref{eq:anomalous_dim_in-in} and \eqref{eq:positivity} are important for understanding the late time behavior of the two point correlation function. The first term \eqref{eq:anomalous_dim_in-in} gives rise to the anomalous dimensions. The second term \eqref{eq:positivity} ensures the positivity of the late time two point correlation function.
For technical reasons, in what follows we will specialize to the case where $\Phi$ is a bulk CFT field with dimension $\delta$.
\subsubsection*{Anomalous dimension}
The anomalous dimension for the external field $\phi$ due to a quadratic interaction with a bulk CFT with dimension $\delta$ can be computed by replacing the two point correlation function of $\phi$ and its commutators inside the nested time integral with their respective leading late time behavior. \eqref{eq:anomalous_dim_in-in} now reduces to 
\begin{align}
\langle\phi(\k,\eta)\phi(\kp,\eta)\rangle'  \supset &-g^2\left\langle\phi(\vec{k}^{\prime}, \eta) \phi(\vec{k}, \eta)\right\rangle_{\Delta_\phi} \frac{1}{2 \nu_\phi} \int_{-\infty}^\eta \mathrm{d} \eta_2 a\left(\eta_2\right)^{d+1}\left(-\eta_2\right)^{\bar{\Delta}_\phi}\notag\\
&\times \int_{-\infty}^{\eta_2} \mathrm{~d} \eta_1 a\left(\eta_1\right)^{d+1}\left(-\eta_1\right)^{\Delta_\phi}
[\Phi_\delta(\vec{k},\eta_1),\Phi_\delta(\vec{k},\eta_2)]\,,
\label{eq:in-in}
\end{align}
where $\Phi_\delta$ is now a CFT operator, as in the main text, and $\langle\phi(\vec{k}^{\prime}, \eta) \phi(\vec{k}, \eta)\rangle_{\Delta_\phi}$ is defined as 
\beq
\langle \phi(\k,\eta)\phi(\kp,\eta)\rangle'_{\Delta_\phi} =\frac{H^2 (-\eta)^3}{2\pi} \Gamma(-i \nu_\phi)^2 \cosh(\tfrac{\pi \nu_\phi}{2} )\left(-\tfrac{k \eta}{2} \right)^{i 2 \nu_\phi} \, .  
\eeq

The first integral over $\eta_1$ can be problematic for certain values of $\delta$. Specifically for $\delta > \frac{d}{2}$, the contact term in \eqref{eq:contact} is divergent at $\eta_1 = \eta_2$ and thus can lead to a divergent integral. This problem can be resolved by computing the integral for $\delta < d/2$ and then analytically continued to all $\delta$. Focusing on the integral over $\eta_1$, we have
\beq
\int_{-\infty}^{\eta_2} \mathrm{~d} \eta_1 a\left(\eta_1\right)^{d+1}\left(-\eta_1\right)^{\Delta_\phi}(-\eta_1)^\delta (\eta_2-\eta_1)^{d-2\delta}
= (-\eta_2)^{\Delta-\delta}\frac{\Gamma(1+d-2\delta)\Gamma(\delta-\Delta_\phi)}{\Gamma(1+d-\delta-\Delta_\phi)}\, , 
\eeq
when $d-2\delta+1 >0$. Despite the fact that the above equality only holds for $\delta <\frac{d+1}{2}$, the RHS of \eqref{eq:eta_integral} is finite for $\delta \geq \frac{d+1}{2}$ and therefore can be used to continue analytically in $\delta$. The nested time integral in \eqref{eq:in-in} now becomes 
\begin{align}
 \langle\phi(\k,\eta)\phi(\kp,\eta)\rangle' & \supset  2^\delta\pi^{d/2+1}\frac{i\lambda^2}{2\nu_\phi}\left\langle\phi(\vec{k}^{\prime}, \eta) \phi(\vec{k}, \eta)\right\rangle_{\Delta_\phi} \notag\\[2pt]
 &\hspace{12pt}   \frac{\Gamma(\delta-\frac{d}{2})\Gamma(\delta-\Delta_\phi)\Gamma(1+d-2\delta)}{\Gamma(\delta)\Gamma(1+d-\delta-\Delta_\phi)}
\sin\left(\pi\left(\frac{d}{2}-\delta\right)\right) \log(-k \eta) \, .
\end{align}
From here, we see the origin of the logarithmic term and the contributions to RG flow.

\section{Flat-space and Euclidean dS Two-point Functions}

\subsection{Flat-space recap}\label{app:2pcf_flat}
To facilitate our understanding of the various integral representation of propagators in dS, it will be useful to briefly recap the equivalent setup in $d+1$-dimensional flat-space for a massive scalar field $\phi$. It is standard to define the momentum space propagator in Euclidean signature, where Fourier space is the natural basis to express a Lorentz invariant two-point function
\begin{equation}
    G(x,y) = \int \frac{\ud^{d+1}k}{(2\pi)^{D}} \tilde{G}(k) e^{ i \mb{k}\cdot (\mb{x}-\mb{y})} 
\end{equation}
where $\tilde{G}(k)$ is the momentum space propagator. In the free theory, $\tilde{G}(k)=1/(k^2+ m^2)$ and its poles at $k_* = \pm i m$ set the infrared scaling, but interactions typically shift the location of this pole. Performing the angular integral leads to the manifestly Lorentz invariant (rotation invariant in Euclidean signature) expression
\begin{equation}
    G(r) = \int_0^{\infty} \frac{\ud k }{(2\pi)^{\frac{d+1}{2}}}\,k^{d} (k r)^{\frac{1-d}{2}} J_{\frac{d-1}{2}}(k r) \tilde{G}(k)\,,
\end{equation}
where $r\equiv |\mb{x}-\mb{y}|$. Now we would like to massage this expression into a form which makes the relationship between the long distance ($r \to +\infty$) asymptotics and the analytic structure of the momentum space propagator $\tilde{G}(k)$ clear, analogous to the Watson-Sommerfeld representation in dS.

By exploiting suitable analytic properties of the Bessel function\footnote{Specifically we can use the fact that $J_{\nu}(z) = \frac{1}{\pi i}\left[e^{-i\pi\nu/2}K_\nu(-i z)-e^{i\pi\nu/2}K_\nu(iz)\right]$, which holds on the right half $z$-plane.} one can check that this integral can be expressed as a contour integral on the complex $k$-plane
\begin{equation}\label{eq:flat_wats_somm}
    G(r) = \int_{-\infty+i \epsilon}^{\infty + i\epsilon} \frac{\ud k}{2\pi i}\, 2k\,\tilde{G}(k)\, G(-ik;r) \,,
\end{equation}
where 
\begin{equation}
    G(m;r) \equiv \frac{1}{(2\pi)^{\frac{d+1}{2}}}\left(\frac{m}{r}\right)^{\frac{d-1}{2}}K_{\frac{d-1}{2}}(mr)
\end{equation}
is the free propagator for a scalar with mass $m$. The form of (\ref{eq:flat_wats_somm}) makes it clear that this is the flat-space analogue of the Watson-Sommerfeld representation in dS.

As we well know, the closest pole we encounter on the ${\rm Im}(k)$ axis as we deform the contour upwards will set the leading decay profile of the propagator, and this is what we commonly understand as the on-shell mass of the particle. For the free propagator $\tilde{G}(k)=1/(k^2+m^2)$, the pole is at $k_*=i m$, accounting for which merely returns the free propagator to us.

In an interacting theory, the pole is shifted away from the free field location and can possess a non-zero imaginary component. Let us say the location of this pole is $m_*$. Accounting for the residue at this point yields
\begin{equation}
    G(t) \propto \frac{1}{m_*}\left(\frac{m_*}{t}\right)^{d/2}e^{-i m_* t} + \cdots.
\end{equation}
The pole $m_*$ may a have both a real and imaginary part, i.e.~$m_*=m - i \gamma$. The real part $m$ sets the oscillation frequency, and this is what we call the physical mass. The imaginary part $\gamma$ induces a decay over long time-scales and is termed the anomalous dimension. The situation in dS is analogous.

The usual Kallen-Lehmann representation can also be obtained using this expression. To do so we can deform the integration contour so that it wraps around the ${\rm Im}(k)$ axis and change variables to $k=i \mu$. We expect either poles or branch cuts along this axis so we will pick up the difference of the momentum space propagator along the line $\mu \in (0,\infty)$. We can therefore define
\begin{equation}
    2\pi \rho(\mu) \equiv 2 i \mu \lim_{\epsilon \to 0^+}\left[\tilde{G}(i\mu+\epsilon)-\tilde{G}(i\mu-\epsilon)\right]\,,
\end{equation}
where $\rho(\mu)$ is the spectral density. With this identification we have
\begin{equation}
    G(r) = \int_0^\infty \ud \mu\, \rho(\mu) G(\mu;r)\,,
\end{equation}
which is the standard flat-space result.
\subsection{Anomalous Dimensions From Euclidean dS}
\label{app:anom_dims_euclidean}
In this section we will quickly review perturbation theory on the Euclidean sphere specifically targeting the calculation of the anomalous dimension. As we shall see, the calculation closely parallels standard methodology in flat space.

A scalar field can be expressed as a sum over the eigenfunctions of the Laplace operator on the sphere, namely spherical harmonics
\begin{equation}
    \phi(x) = \sum_{\mb{J}} \varphi_{\mb{J}} Y_{\mb{J}}(x)\,,
\end{equation}
where $\mb{J}\equiv (J, \mb{m})$ is the vector composed of the total angular momentum and the azimuthal angular momenta respectively. The dS group upon performing this Wick rotation is merely the rotation group $SO(d+1)$. The two-point function, constrained by rotation invariance, can be expressed a sum over Gegenbauer polynomials
\begin{equation}
    \langle \phi(x) \phi(y)\rangle = \frac{\Gamma(\tfrac{d}{2})}{2\pi^{\tfrac{d+2}{2}}}\sum_{J=0}^{\infty}(J+\tfrac{d}{2})\momcoeff{\phi}{J} C_J^{\tfrac{d}{2}}(\xi) \,,
\end{equation}
where $\xi\equiv \vec{x}\cdot \vec{y}$ is the embedding distance on the sphere. The term $\momcoeff{\phi}{J}$ is called the momentum coefficient although, as discussed in the main text, it bears a close resemblance to the Lorentz invariant momentum propagator in flat-space. 

Our goal is to compute loop corrections to the two-point function of $\phi$, where we will restrict to diagrams of the bubble topology. A complication of working in Euclidean signature is that enforcing momentum conservation on vertices is non-trivial. In flat-space, and spatial momentum space, translation invariance guarantees $\sum_i k_i=0$ at any vertex. On the sphere, it need not be the case that $\sum_i \mb{J}_i=0$, as is familiar from the representation theory of the rotation group in two-dimensions. However, by restricting to bubble diagrams, we essentially restrict our attention to bi-linear interactions for which momentum conservation is identical to flat space. 

Let us see this in more detail. We are interested in the general theory 
\begin{equation}
    S_{\rm E} = \int \ud^{d+1}x\,\sqrt{g}\left[\tfrac{1}{2} (\partial \phi)^2 + \tfrac{1}{2}m_\phi^2 \phi^2 + g\s \phi\s \Phi\right] + S_{\rm ct}\,,
\end{equation}
where $\Phi$ is any scalar operator, the correlators of which we will comment on briefly. The counterterm action is
\begin{equation}
    S_{\rm ct} = \int\ud^{d+1}x\,\sqrt{g} \left[\tfrac{1}{2}\delta_{Z_\phi} (\partial \phi)^2 + \tfrac{1}{2}\delta_{m_\phi}\phi^2\right]\,,
\end{equation}
where we will not require a counterterm for the coupling constant since we are only interested in the leading order corrections to the two-point function. The momentum space propagators can be read off from the free action
\begin{equation}
\def\ylvl{0}
    \begin{tikzpicture}[thick, baseline=-3pt]
        \coordinate (c1) at (-1., \ylvl);
        \coordinate (c3) at (1., \ylvl);
        \draw[black] (c1) -- (c3) ;
        \end{tikzpicture} = \momcoeff{\phi}{J}\,\qquad \text{and}\,\qquad 
        \begin{tikzpicture}[thick, baseline=-3pt]
        \coordinate (c1) at (-1., \ylvl);
        \coordinate (c3) at (1., \ylvl);
        \draw[pyRed, double] (c1) -- (c3) ;
        \end{tikzpicture} = \momcoeff{\Phi}{J}\,.
\end{equation}
The interaction vertices are
\begin{equation}
\def\ylvl{0}
\def\circSize{0.1}
    \begin{aligned}
        \begin{tikzpicture}[thick, baseline=-3pt]
        \coordinate (c1) at (-1., \ylvl);
        \coordinate (c2) at (0, \ylvl);
        \coordinate (c3) at (1., \ylvl);
        \begin{scope}[shift={(0, \ylvl)}]
            \draw[pyRed, double] (c1) -- (c2);
        \end{scope}

        \draw[black] (c2) -- (c3) ;

        \fill[intSty] (c2) circle (0.07) ;
        \end{tikzpicture} &= -g\qquad\text{and}\qquad \begin{tikzpicture}[thick, baseline=-3pt]
        \coordinate (c1) at (-1., \ylvl);
        \coordinate (c2) at (0, \ylvl);
        \coordinate (c3) at (1., \ylvl);
        \draw[black] (c1) -- (c2) ;

        \draw[black] (c2) -- (c3) ;

        \fill[white, draw=black, line width=0.3mm] (0, \ylvl) circle (\circSize);
        \draw[rotate=45, line width=0.25mm] (-\circSize, 0) -- (\circSize, 0);
        \draw[rotate=-45, line width=0.25mm] (-\circSize, 0) -- (\circSize, 0);
        \end{tikzpicture} = -J(J+d) \delta_{Z_\phi} -\delta_{m_\phi}.
    \end{aligned}
\end{equation}
The momentum space propagator of $\phi$ will receive corrections due to the interaction with $\Phi$. When $\Phi$ is composed of polynomials of a free field, e.g.~$\sigma^n$, the correlators of $\Phi$ can be Wick factorized into a sum over disconnected contributions. This straightforwardly enables the existence of a diagram at $\mathcal{O}(g^{2k})$ which takes the form of a chain of bubble diagrams
\begin{equation}
\def\ylvl{0}
\def\circSizeBub{0.45}
\def\circSize{0.1}
    \momcoeff{\phi}{J} = \begin{tikzpicture}[thick, baseline=-3pt]
        \coordinate (c1) at (-1., \ylvl);
        \coordinate (c3) at (1., \ylvl);
        \draw[black] (c1) -- (c3) ;
        \end{tikzpicture} + \begin{tikzpicture}[thick, baseline=-3pt]
        \coordinate (c1) at (-1., \ylvl);
        \coordinate (c3) at (1., \ylvl);
        \draw[black] (c1) -- (c3) ;
        \fill[white, draw=black, line width=0.3mm] (0, \ylvl) circle (\circSizeBub);
        \begin{scope}
            \clip[draw] (0, 0) circle (\circSizeBub);
            \foreach \x in {-0.5, -0.425, ..., 0.5} 
            {	
                \draw[rotate=45, thin] (-\circSizeBub, \x) -- (\circSizeBub, \x);
            }
        \end{scope}
        \end{tikzpicture} + 
        \begin{tikzpicture}[thick, baseline=-3pt]
        \coordinate (c1) at (-1.75, \ylvl);
        \coordinate (c3) at (1.75, \ylvl);
        \draw[black] (c1) -- (c3) ;
        \fill[white, draw=black, line width=0.3mm] (-0.75, \ylvl) circle (\circSizeBub);
        \fill[white, draw=black, line width=0.3mm] (0.75, \ylvl) circle (\circSizeBub);
        \begin{scope}[shift={(-0.75, 0)}]
            \clip[draw] (0, 0) circle (\circSizeBub);
            \foreach \x in {-0.5, -0.425, ..., 0.5} 
            {	
                \draw[rotate=45, thin] (-\circSizeBub, \x) -- (\circSizeBub, \x);
            }
        \end{scope}
        \begin{scope}[shift={(0.75, 0)}]
            \clip[draw] (0, 0) circle (\circSizeBub);
            \foreach \x in {-0.5, -0.425, ..., 0.5} 
            {	
                \draw[rotate=45, thin] (-\circSizeBub, \x) -- (\circSizeBub, \x);
            }
        \end{scope}
        \end{tikzpicture} + \cdots\,,
\end{equation}
which develops a hierarchy of singularities as we approach the the free-field poles $J\to -\Delta_\phi,-\bar{\Delta}_\phi$ which corresponds to sending a tower of $\phi$ states on-shell. The sequence of bubble diagrams must be 1PI resummed, in exact analogy with flat-space perturbative computations, which enables a notion of self-energy. Notably, the main operator of interest in this paper, the vertex operator, does not Wick factorize. Nevertheless, it was shown in \cite{Chakraborty:2023eoq} that the resummation is still necessary in order to analyze perturbative corrections in this case owing to an effective Wick factorization of the vertex operators at long distances.

Using the Feynman rules it is straightforward to determine the leading order corrections to the $\phi$ propagator
\begin{equation}
    \def\ylvl{0}
    \def\circSizeBub{0.45}
    \def\circSize{0.1}
    \begin{tikzpicture}[thick, baseline=-3pt]
        \coordinate (c1) at (-1., \ylvl);
        \coordinate (c3) at (1., \ylvl);
        \draw[black] (c1) -- (c3) ;
        \fill[white, draw=black, line width=0.3mm] (0, \ylvl) circle (\circSizeBub);
        \begin{scope}
            \clip[draw] (0, 0) circle (\circSizeBub);
            \foreach \x in {-0.5, -0.425, ..., 0.5} 
            {	
                \draw[rotate=45, thin] (-\circSizeBub, \x) -- (\circSizeBub, \x);
            }
        \end{scope}
        \end{tikzpicture} = \begin{tikzpicture}[thick, baseline=-3pt]
        \coordinate (c1) at (-1., \ylvl);
        \coordinate (c21) at (-0.45, \ylvl);
        \coordinate (c22) at (0.45, \ylvl);
        \coordinate (c3) at (1., \ylvl);

        \draw[black] (c1) -- (c21) ;
        \draw[pyRed, double] (c21) -- (c22);
        \draw[black] (c22) -- (c3) ;

        \fill[intSty] (c21) circle (0.07) ;
        \fill[intSty] (c22) circle (0.07) ;
        \end{tikzpicture} 
        +
        \begin{tikzpicture}[thick, baseline=-3pt]
        \coordinate (c1) at (-1., \ylvl);
        \coordinate (c2) at (0, \ylvl);
        \coordinate (c3) at (1., \ylvl);
        \draw[black] (c1) -- (c2) ;

        \draw[black] (c2) -- (c3) ;

        \fill[white, draw=black, line width=0.3mm] (0, \ylvl) circle (\circSize);
        \draw[rotate=45, line width=0.25mm] (-\circSize, 0) -- (\circSize, 0);
        \draw[rotate=-45, line width=0.25mm] (-\circSize, 0) -- (\circSize, 0);
        \end{tikzpicture}\,,
\end{equation}
which translates to the self-energy
\begin{equation}\label{eq:self_E}
    \Pi_\phi(J) = g^2 \momcoeff{\Phi}{J} - J(J+d)\delta_{Z_\phi} - \delta_{m_\phi}.
\end{equation}
The quantum corrected momentum coefficient of $\phi$ is then
\begin{equation}\label{eq:phi_momcoeff_corr}
    \momcoeff{\phi}{J} = \frac{1}{(J+\Delta_\phi)(J+\bar{\Delta}_\phi)-\Pi_\phi(J)}.
\end{equation}
The free-field poles $-\Delta_\phi$ and $-\bar{\Delta}_\phi$ will shift due to interactions to $J_*=-\Delta_\phi-\delta J$ and $\bar{J}_*=-\bar{\Delta}_\phi-\delta\bar{J}$, where
\begin{equation}
    \delta J \simeq \frac{\Pi_\phi(-\Delta_\phi)}{\Delta_\phi - \bar{\Delta}_\phi} \qquad {\rm and} \qquad \delta\bar{J} \simeq  -\frac{\Pi_\phi(-\bar{\Delta}_\phi)}{\Delta_\phi - \bar{\Delta}_\phi} \, .
\end{equation}
Here $\delta \bar J$ should be understood as small correction to the location of the pole at $\bar J= - \bar \Delta_\phi$, and not the shadow transform of $\delta J$ (in exactly the same way that $\bar \gamma_\phi$ is a correction to $\bar \Delta_\phi$). These equalities are understood to hold perturbatively, which for our case means to leading order in the coupling, which is ${\cal O}(g^2)$. 

A precise understanding of the locations of the poles depends depends on the counter-terms in \eqref{eq:self_E}. Yet, we observe that the sum of these two poles is independent of the renormalization scheme and can be expressed as
\begin{equation}
    \frac{\delta J + \delta\bar{J}}{2} = \frac{g^2}{2\nu_\phi}{\rm Im}\left(\momcoeff{\Phi}{-\Delta_\phi}\right) \, .
\end{equation}
The difference, on the other hand, is scheme-dependent, and constitutes a definition of the mass. We will implementing an on-shell scheme which requires that the difference between these two poles is zero. We also need to make a choice regarding the normalization of the infrared states, which is accomplished by fixing the residues at $J_*$ and $\bar{J}_*$. Let us call these residues $\mathcal{R}$ and $\bar{\mathcal{R}}$, which can be evaluated to yield
\begin{equation}
    \mathcal{R}^{-1} = -2i\nu_\phi -2 \delta J - \Pi_\phi'(-\Delta_\phi)\quad {\rm and} \quad \bar{\mathcal{R}}^{-1} = 2i\nu_\phi -2 \delta \bar{J} - \Pi_\phi'(-\bar{\Delta}_\phi)\, ,
\end{equation}
where the $'$ denotes a derivative with respect to $J$. Note once again that the sum of these quantities is unambiguous, 
\begin{equation}
    \mathcal{R}^{-1}+\bar{\mathcal{R}}^{-1} =-\frac{2g^2}{\nu_\phi}\left[{\rm Im}\left(\momcoeff{\Phi}{-\Delta_\phi}\right) + \nu_\phi {\rm Re}\left(\momcoeff{\Phi}{-\Delta_\phi}'\right)\right] \, .
\end{equation}
We will prove in \cref{app:dispersive} that the real part of the derivative of the momentum coefficient is UV finite, as required for this sum to be scheme-independent.

Unlike the sum, the difference between the two residues is scheme dependent. We will pick a scheme in which this difference, or the imaginary part, is equal to the free field values $\mathcal{R}^{-1}-\bar{\mathcal{R}}^{-1}=-4i \nu_\phi$. This ensures the residues are complex conjugates of one another. Here we note that in the free theory, the poles were related both by complex conjugation and the shadow transform. The the location of the shifted poles are not related by the shadow transform, but are still related by complex conjugation, which suggests we should maintain this property of the residues as well. Implement this condition has some non-trivial consequences because the counter-terms are explicitly shadow symmetric. Imposing the renormalization conditions on the difference of residues and the difference of the shifted poles ($\delta J=\delta \bar{J}$) determines $\delta_{Z_\phi}$ and $ \delta_{m_\phi} $ respectively, 
\begin{subequations}
    \begin{align}
        \delta_{Z_\phi}  &= - \frac{g^2}{2\nu_\phi}{\rm Im}\left(\momcoeff{\Phi}{-\Delta_\phi}'\right) \\
        \delta_{m_\phi} &= g^2 {\rm Re}\left(\momcoeff{\Phi}{-\Delta_\phi}\right) - \frac{g^2}{2\nu_\phi}\Delta_\phi \bar{\Delta}_\phi {\rm Im}\left(\momcoeff{\Phi}{-\Delta_\phi}'\right) \, .
    \end{align}
\end{subequations}
We will see in the next subsection that both quantities are UV divergent. The shift of the poles define the anomalous dimension $\delta J=\delta \bar{J}\equiv \gamma_\phi$, which takes the value
\begin{equation}
    \gamma_\phi = \frac{g^2}{2\nu_\phi}{\rm Im}(\momcoeff{\Phi}{-\Delta_\phi}) = \frac{g^2}{4\nu_\phi^2} \rho_\Phi(\Delta_\phi)\,.
\end{equation}
The residue can also be determined to be
\begin{align}
        \mathcal{R}^{-1} &= -2i\nu_\phi -\frac{g^2}{\nu_\phi}\left[{\rm Im}\left(\momcoeff{\Phi}{-\Delta_\phi}\right) + \nu_\phi {\rm Re}\left(\momcoeff{\Phi}{-\Delta_\phi}'\right)\right]\notag\\
        &=-2i\nu_\phi -2\gamma_\phi + i \frac{g^2}{\nu_\phi}\rho_\Phi'(\Delta_\phi)\,,
\end{align}
which demonstrates that a non-zero anomalous dimension can result in a change in the normalization of the infrared state. Note that the last term on the second line is real valued since $\rho'(\Delta_\phi)$ is purely imaginary. To obtain this expression we have related the momentum coefficient and its derivative to the spectral density and its derivative through (\ref{eq:momcoefftospectral}) and (\ref{eq:momderivdisp}).

\begin{figure}[t!]
    \centering
    \includegraphics[width=0.7\linewidth]{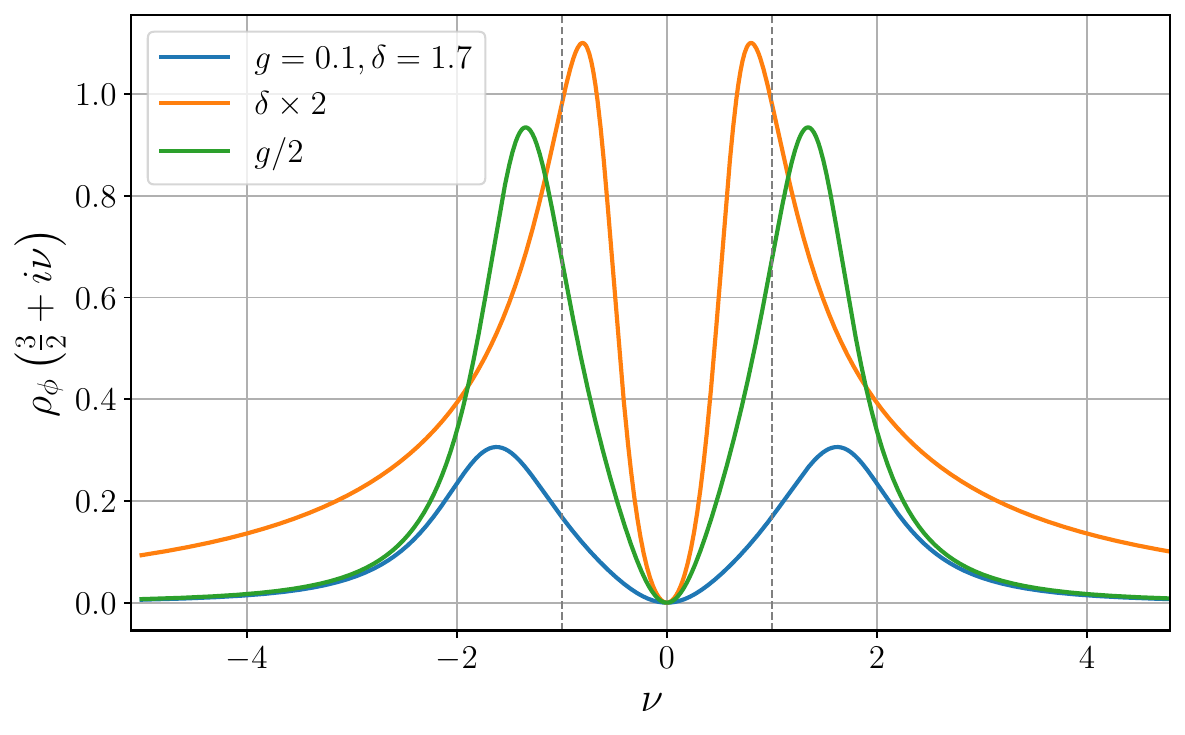}
    \caption{We plot the leading order corrected spectral density for a principal series scalar with $\nu_\phi=1$ coupled to a CFT operator with scaling dimension $\delta$, for various choices of coupling $g$ and scaling dimension. The dashed vertical lines indicate $\nu=\pm \nu_\phi$.}
    \label{fig:example_corr_spectral_density}
\end{figure}

Finally, we can determine the spectral density $\rho_\phi(\Delta)$ in the interacting theory using the momentum coefficient (\ref{eq:phi_momcoeff_corr}) by extracting the imaginary part (see also e.g.~\cite{Loparco:2025azm})
\begin{equation}
    \rho_\phi(\Delta) = \frac{g^2 \rho_\Phi(\Delta)}{\left[(\Delta-\Delta_\phi)(\Delta-\bar{\Delta}_\phi)-\Pi_\phi(-\Delta)\right]\left[(\bar{\Delta}-\Delta_\phi)(\bar{\Delta}-\bar{\Delta}_\phi)-\Pi_\phi(-\bar{\Delta})\right]}.
\end{equation}
As an example, we show in Figure~\ref{fig:example_corr_spectral_density} the corrected spectral density taking $\Phi$ to be a CFT operator with scaling dimension $\delta$, following our renormalization procedure. 

\subsection{Dispersive Identities and the Running Mass}\label{app:dispersive}

Now we will briefly discuss the regularization of the bubble diagram and also comment on the running of the mass. To do so we will rely on certain dispersion relations between the spectral density. In order to regularize the self-energy we will truncate the set of states, which practically corresponds to a cutoff in the KL integral at a maximum mass $\bar{\nu}$. This procedure is analogous to previous implementations in AdS \cite{Fitzpatrick:2010zm}.\footnote{Another convenient dS invariant choice is dimensional regularization via analytic continuation in $d$ \cite{Marolf:2010zp}.}

The identity (\ref{eq:momcoefftospectral}) can be inverted and the momentum coefficient can be expressed as a dispersive integral~\cite{Chakraborty:2025myb,Loparco:2025azm}
\begin{equation}\label{eq:momcoeffdispersive}
    \momcoeff{\Phi}{J} = \int_{\mathcal{C}} \frac{\ud \Delta}{2\pi i} \frac{\rho_\Phi(\Delta)}{(J+\Delta)(J+\bar{\Delta})}\,,
\end{equation}
where the contour $\mathcal{C}$ runs along the principal series axis and ${\rm Re}(J)> -\frac{d}{2}$. Taking the imaginary part of the integral straightforwardly yields
\begin{equation}
    {\rm Im}\left(\momcoeff{\Phi}{-\Delta_\phi}\right) = \frac{1}{2\nu_\phi}\rho_\Phi(\Delta_\phi) \, .
\end{equation}
Let us now analyze the derivative. Differentiating (\ref{eq:momcoeffdispersive}) yields
\begin{equation}
    \momcoeff{\Phi}{J}' = -(d+2J)\int_{\mathcal{C}} \frac{\ud \Delta}{2\pi i} \frac{\rho_\Phi(\Delta)}{(J+\Delta)^2(J+\bar{\Delta})^2}.
\end{equation}
Using this we can determine the sum
\begin{align}
        \momcoeff{\Phi}{-\Delta_\phi}' + \momcoeff{\Phi}{-\bar{\Delta}_\phi}' &= (2\Delta_\phi-d) \left[{\rm Res}_{\Delta_\phi}-{\rm Res}_{\bar{\Delta}_\phi}\right]\frac{\rho_\Phi(\Delta)}{(J+\Delta)^2(J+\bar{\Delta})^2} \notag\\
        &= -\frac{2}{(2\Delta_\phi-d)^2}\left[2 \rho_\Phi(\Delta_\phi)+(2\Delta_\phi-d)\rho_\Phi'(\Delta_\phi)\right]\,,
        \label{eq:momderivdisp}
\end{align}
where we have additionally simplified the expression by noting that $\rho_\Phi'(\Delta)=-\rho_\Phi'(d-\Delta)$, or the derivative of a symmetric function is antisymmetric. Keeping in mind that the sum on the left hand side is simply the real part of $\momcoeff{\Phi}{-\Delta_\phi}$, this analysis proves that it is indeed UV finite. 

The real part of the momentum coefficient and the imaginary part of the derivative are sensitive to the full integral over $\Delta$, and thus can be UV divergent. Let us use this fact to determine the running of the mass, which as we will see, is fairly straightforward to determine. To do so we return to (\ref{eq:momcoeffdispersive}) and change variables to $\nu = -i(\Delta-\tfrac{d}{2})$ and regulate the integral by truncating it at the cutoff $\Lambda \equiv \bar{\nu} H$, whereupon we restore factors of Hubble. Let us begin with the momentum coefficient 
\begin{equation}
    \momcoeff{\Phi}{-\Delta_\phi} = \int_{-\bar{\nu}}^{\bar{\nu}} \frac{\ud \nu}{2\pi} \frac{\rho_\Phi(\tfrac{d}{2}+i \nu)}{\nu^2-\nu_\phi^2}\,,
\end{equation}
which upon differentiating yields
\begin{equation}
    H\partial_\Lambda\momcoeff{\Phi}{-\Delta_\phi} = \frac{H^2}{\pi \Lambda^2} \rho_\Phi(\tfrac{d}{2}+i \bar{\nu})\,,
\end{equation}
which we see is real valued, as promised. Note that we have approximated $\bar{\nu} \gg \nu_\phi$. We can do the same with the derivative
\begin{equation}
    \momcoeff{\Phi}{-\Delta_\phi}' = 2i \nu_\phi \int_{-\bar{\nu}}^{\bar{\nu}} \frac{\ud \nu}{2\pi} \frac{\rho_\Phi(\tfrac{d}{2}+i \nu)}{(\nu^2-\nu_\phi^2)^2}\,,
\end{equation}
which upon differentiating yields
\begin{equation}
    H\partial_\Lambda\momcoeff{\Phi}{-\Delta_\phi}' = 2i\frac{\nu_\phi H^4}{\pi \Lambda^4} \rho_\Phi(\tfrac{d}{2}+i \bar{\nu})\,,
\end{equation}
which is purely imaginary. Using these results we can determine the running of the mass
\begin{equation}
    \Lambda \partial_\Lambda\delta_{m_\phi} \simeq H^{-2}g^2\frac{H}{\pi \Lambda} \rho_\Phi(\tfrac{d}{2}+i \bar{\nu})\,,
\end{equation}
where we have neglected the running of the derivative, which is subleading in $\Lambda$.

\clearpage
\phantomsection
\addcontentsline{toc}{section}{References}
\small
\bibliographystyle{utphys}
\bibliography{Refs}

@article{Salcedo:2024smn,
    author = "Salcedo, Santiago Agui and Colas, Thomas and Pajer, Enrico",
    title = "{The open effective field theory of inflation}",
    eprint = "2404.15416",
    archivePrefix = "arXiv",
    primaryClass = "hep-th",
    doi = "10.1007/JHEP10(2024)248",
    journal = "JHEP",
    volume = "10",
    pages = "248",
    year = "2024"
}

@article{Green:2024cmx,
    author = "Green, Daniel and Sun, Guanhao",
    title = "{Effective field theory and in-in correlators}",
    eprint = "2412.02739",
    archivePrefix = "arXiv",
    primaryClass = "hep-th",
    doi = "10.1007/JHEP04(2025)166",
    journal = "JHEP",
    volume = "04",
    pages = "166",
    year = "2025"
}

@article{Colas:2025ind,
    author = "Colas, Thomas and Qin, Zhehan and Tong, Xi",
    title = "{Open Effective Field Theory and the Physics of Cosmological Collider Signals}",
    eprint = "2512.07941",
    archivePrefix = "arXiv",
    primaryClass = "hep-th",
    month = "12",
    year = "2025"
}

@article{Green:2025hmo,
    author = "Green, Daniel and Gupta, Kshitij",
    title = "{Quantum Walks and Exact RG in de Sitter Space}",
    eprint = "2512.13842",
    archivePrefix = "arXiv",
    primaryClass = "hep-th",
    month = "12",
    year = "2025"
}

@article{Green:2023ids,
    author = "Green, Daniel and Huang, Yiwen and Shen, Chia-Hsien and Baumann, Daniel",
    title = "{Positivity from Cosmological Correlators}",
    eprint = "2310.02490",
    archivePrefix = "arXiv",
    primaryClass = "hep-th",
    doi = "10.1007/JHEP04(2024)034",
    journal = "JHEP",
    volume = "04",
    pages = "034",
    year = "2024"
}

@article{Fitzpatrick:2010zm,
    author = "Fitzpatrick, A. Liam and Katz, Emanuel and Poland, David and Simmons-Duffin, David",
    title = "{Effective Conformal Theory and the Flat-Space Limit of AdS}",
    eprint = "1007.2412",
    archivePrefix = "arXiv",
    primaryClass = "hep-th",
    reportNumber = "BUHET-07-14-10",
    doi = "10.1007/JHEP07(2011)023",
    journal = "JHEP",
    volume = "07",
    pages = "023",
    year = "2011"
}

@article{Strominger:2001pn,
    author = "Strominger, Andrew",
    title = "{The dS / CFT correspondence}",
    eprint = "hep-th/0106113",
    archivePrefix = "arXiv",
    doi = "10.1088/1126-6708/2001/10/034",
    journal = "JHEP",
    volume = "10",
    pages = "034",
    year = "2001"
}

@article{Creminelli:2011mw,
    author = "Creminelli, Paolo",
    title = "{Conformal invariance of scalar perturbations in inflation}",
    eprint = "1108.0874",
    archivePrefix = "arXiv",
    primaryClass = "hep-th",
    doi = "10.1103/PhysRevD.85.041302",
    journal = "Phys. Rev. D",
    volume = "85",
    pages = "041302",
    year = "2012"
}

@article{Maldacena:2011nz,
    author = "Maldacena, Juan M. and Pimentel, Guilherme L.",
    title = "{On graviton non-Gaussianities during inflation}",
    eprint = "1104.2846",
    archivePrefix = "arXiv",
    primaryClass = "hep-th",
    reportNumber = "PUPT-2371",
    doi = "10.1007/JHEP09(2011)045",
    journal = "JHEP",
    volume = "09",
    pages = "045",
    year = "2011"
}

@article{Gleyzes:2016tdh,
    author = "Gleyzes, J{\'e}r{\^o}me and de Putter, Roland and Green, Daniel and Dor{\'e}, Olivier",
    title = "{Biasing and the search for primordial non-Gaussianity beyond the local type}",
    eprint = "1612.06366",
    archivePrefix = "arXiv",
    primaryClass = "astro-ph.CO",
    doi = "10.1088/1475-7516/2017/04/002",
    journal = "JCAP",
    volume = "04",
    pages = "002",
    year = "2017"
}

@article{Lee:2016vti,
    author = "Lee, Hayden and Baumann, Daniel and Pimentel, Guilherme L.",
    title = "{Non-Gaussianity as a Particle Detector}",
    eprint = "1607.03735",
    archivePrefix = "arXiv",
    primaryClass = "hep-th",
    doi = "10.1007/JHEP12(2016)040",
    journal = "JHEP",
    volume = "12",
    pages = "040",
    year = "2016"
}

@article{Chen:2012ge,
    author = "Chen, Xingang and Wang, Yi",
    title = "{Quasi-Single Field Inflation with Large Mass}",
    eprint = "1205.0160",
    archivePrefix = "arXiv",
    primaryClass = "hep-th",
    doi = "10.1088/1475-7516/2012/09/021",
    journal = "JCAP",
    volume = "09",
    pages = "021",
    year = "2012"
}

@article{Sohn:2024xzd,
    author = "Sohn, Wuhyun and Wang, Dong-Gang and Fergusson, James R. and Shellard, E. P. S.",
    title = "{Searching for cosmological collider in the Planck CMB data}",
    eprint = "2404.07203",
    archivePrefix = "arXiv",
    primaryClass = "astro-ph.CO",
    doi = "10.1088/1475-7516/2024/09/016",
    journal = "JCAP",
    volume = "09",
    pages = "016",
    year = "2024"
}

@article{Jazayeri:2023xcj,
    author = "Jazayeri, Sadra and Renaux-Petel, S{\'e}bastien and Werth, Denis",
    title = "{Shapes of the cosmological low-speed collider}",
    eprint = "2307.01751",
    archivePrefix = "arXiv",
    primaryClass = "hep-th",
    doi = "10.1088/1475-7516/2023/12/035",
    journal = "JCAP",
    volume = "12",
    pages = "035",
    year = "2023"
}

@article{Meltzer:2021zin,
    author = "Meltzer, David",
    title = "{The inflationary wavefunction from analyticity and factorization}",
    eprint = "2107.10266",
    archivePrefix = "arXiv",
    primaryClass = "hep-th",
    reportNumber = "CALT-TH-2021-028",
    doi = "10.1088/1475-7516/2021/12/018",
    journal = "JCAP",
    volume = "12",
    number = "12",
    pages = "018",
    year = "2021"
}

@article{Goodhew:2024eup,
    author = "Goodhew, Harry and Thavanesan, Ayngaran and Wall, Aron C.",
    title = "{The Cosmological CPT Theorem}",
    eprint = "2408.17406",
    archivePrefix = "arXiv",
    primaryClass = "hep-th",
    month = "8",
    year = "2024"
}

@article{Pueyo:2024twm,
    author = "Pueyo, Carlos Duaso and Goodhew, Harry and McCulloch, Ciaran and Pajer, Enrico",
    title = "{Perturbative unitarity bounds from momentum-space entanglement}",
    eprint = "2410.23709",
    archivePrefix = "arXiv",
    primaryClass = "hep-th",
    month = "10",
    year = "2024"
}

@article{Melville:2024ove,
    author = "Melville, Scott and Pimentel, Guilherme L.",
    title = "{A de Sitter S-matrix from amputated cosmological correlators}",
    eprint = "2404.05712",
    archivePrefix = "arXiv",
    primaryClass = "hep-th",
    doi = "10.1007/JHEP08(2024)211",
    journal = "JHEP",
    volume = "08",
    pages = "211",
    year = "2024"
}

@article{Goodhew:2021oqg,
    author = "Goodhew, Harry and Jazayeri, Sadra and Lee, Mang Hei Gordon and Pajer, Enrico",
    title = "{Cutting cosmological correlators}",
    eprint = "2104.06587",
    archivePrefix = "arXiv",
    primaryClass = "hep-th",
    doi = "10.1088/1475-7516/2021/08/003",
    journal = "JCAP",
    volume = "08",
    pages = "003",
    year = "2021"
}

@article{Melville:2021lst,
    author = "Melville, Scott and Pajer, Enrico",
    title = "{Cosmological Cutting Rules}",
    eprint = "2103.09832",
    archivePrefix = "arXiv",
    primaryClass = "hep-th",
    doi = "10.1007/JHEP05(2021)249",
    journal = "JHEP",
    volume = "05",
    pages = "249",
    year = "2021"
}

@article{Kumar:2019ebj,
    author = "Kumar, Soubhik and Sundrum, Raman",
    title = "{Cosmological Collider Physics and the Curvaton}",
    eprint = "1908.11378",
    archivePrefix = "arXiv",
    primaryClass = "hep-ph",
    reportNumber = "UMD-PP-019-04",
    doi = "10.1007/JHEP04(2020)077",
    journal = "JHEP",
    volume = "04",
    pages = "077",
    year = "2020"
}

@article{Chen:2016uwp,
    author = "Chen, Xingang and Wang, Yi and Xianyu, Zhong-Zhi",
    title = "{Standard Model Background of the Cosmological Collider}",
    eprint = "1610.06597",
    archivePrefix = "arXiv",
    primaryClass = "hep-th",
    doi = "10.1103/PhysRevLett.118.261302",
    journal = "Phys. Rev. Lett.",
    volume = "118",
    number = "26",
    pages = "261302",
    year = "2017"
}

@article{Meerburg:2016zdz,
    author = {Meerburg, P. Daniel and M{\"u}nchmeyer, Moritz and Mu{\~n}oz, Julian B. and Chen, Xingang},
    title = "{Prospects for Cosmological Collider Physics}",
    eprint = "1610.06559",
    archivePrefix = "arXiv",
    primaryClass = "astro-ph.CO",
    doi = "10.1088/1475-7516/2017/03/050",
    journal = "JCAP",
    volume = "03",
    pages = "050",
    year = "2017"
}

@article{Green:2024hbw,
    author = "Green, Daniel and Gupta, Kshitij and Huang, Yiwen",
    title = "{A Goldstone boson equivalence for inflation}",
    eprint = "2403.05274",
    archivePrefix = "arXiv",
    primaryClass = "hep-th",
    doi = "10.1007/JHEP09(2024)117",
    journal = "JHEP",
    volume = "09",
    pages = "117",
    year = "2024"
}

@article{Green:2024fsz,
    author = "Green, Daniel and Gupta, Kshitij",
    title = "{Soft Metric Fluctuations During Inflation}",
    eprint = "2410.11973",
    archivePrefix = "arXiv",
    primaryClass = "hep-th",
    month = "10",
    year = "2024"
}

@article{Flauger:2009ab,
    author = "Flauger, Raphael and McAllister, Liam and Pajer, Enrico and Westphal, Alexander and Xu, Gang",
    title = "{Oscillations in the CMB from Axion Monodromy Inflation}",
    eprint = "0907.2916",
    archivePrefix = "arXiv",
    primaryClass = "hep-th",
    reportNumber = "SLAC-PUB-14821",
    doi = "10.1088/1475-7516/2010/06/009",
    journal = "JCAP",
    volume = "06",
    pages = "009",
    year = "2010"
}

@article{Chen:2006nt,
    author = "Chen, Xingang and Huang, Min-xin and Kachru, Shamit and Shiu, Gary",
    title = "{Observational signatures and non-Gaussianities of general single field inflation}",
    eprint = "hep-th/0605045",
    archivePrefix = "arXiv",
    reportNumber = "SLAC-PUB-11840, MAD-TH-06-3, UFIFT-HEP-06-9, SU-ITP-06-12, CU-TP-1147",
    doi = "10.1088/1475-7516/2007/01/002",
    journal = "JCAP",
    volume = "01",
    pages = "002",
    year = "2007"
}

@article{Cheung:2007st,
    author = "Cheung, Clifford and Creminelli, Paolo and Fitzpatrick, A. Liam and Kaplan, Jared and Senatore, Leonardo",
    title = "{The Effective Field Theory of Inflation}",
    eprint = "0709.0293",
    archivePrefix = "arXiv",
    primaryClass = "hep-th",
    reportNumber = "IC-2007-032",
    doi = "10.1088/1126-6708/2008/03/014",
    journal = "JHEP",
    volume = "03",
    pages = "014",
    year = "2008"
}

@article{Babich:2004gb,
    author = "Babich, Daniel and Creminelli, Paolo and Zaldarriaga, Matias",
    title = "{The Shape of non-Gaussianities}",
    eprint = "astro-ph/0405356",
    archivePrefix = "arXiv",
    reportNumber = "HUTP-04-A022",
    doi = "10.1088/1475-7516/2004/08/009",
    journal = "JCAP",
    volume = "08",
    pages = "009",
    year = "2004"
}

@article{Planck:2019kim,
    author = "Akrami, Y. and others",
    collaboration = "Planck",
    title = "{Planck 2018 results. IX. Constraints on primordial non-Gaussianity}",
    eprint = "1905.05697",
    archivePrefix = "arXiv",
    primaryClass = "astro-ph.CO",
    doi = "10.1051/0004-6361/201935891",
    journal = "Astron. Astrophys.",
    volume = "641",
    pages = "A9",
    year = "2020"
}

@inproceedings{Poland:2022qrs,
    author = "Poland, David and Simmons-Duffin, David",
    title = "{Snowmass White Paper: The Numerical Conformal Bootstrap}",
    booktitle = "{Snowmass 2021}",
    eprint = "2203.08117",
    archivePrefix = "arXiv",
    primaryClass = "hep-th",
    reportNumber = "CALT-TH 2022-013",
    month = "3",
    year = "2022"
}

@article{Antoniadis:2011ib,
    author = "Antoniadis, Ignatios and Mazur, Pawel O. and Mottola, Emil",
    title = "{Conformal Invariance, Dark Energy, and CMB Non-Gaussianity}",
    eprint = "1103.4164",
    archivePrefix = "arXiv",
    primaryClass = "gr-qc",
    reportNumber = "LA-UR-11-10115, CERN-PH-TH-2011-057",
    doi = "10.1088/1475-7516/2012/09/024",
    journal = "JCAP",
    volume = "09",
    pages = "024",
    year = "2012"
}

@inproceedings{Spradlin:2001pw,
    author = "Spradlin, Marcus and Strominger, Andrew and Volovich, Anastasia",
    title = "{Les Houches lectures on de Sitter space}",
    booktitle = "{Les Houches Summer School: Session 76: Euro Summer School on Unity of Fundamental Physics: Gravity, Gauge Theory and Strings}",
    eprint = "hep-th/0110007",
    archivePrefix = "arXiv",
    pages = "423--453",
    month = "10",
    year = "2001"
}

@article{dsrep0,
 ISSN = {00804630},
 URL = {http://www.jstor.org/stable/97833},
 author = {Harish-Chandra},
 journal = {Proceedings of the Royal Society of London. Series A, Mathematical and Physical Sciences},
 number = {1018},
 pages = {372--401},
 publisher = {The Royal Society},
 title = {Infinite Irreducible Representations of the Lorentz Group},
 urldate = {2024-06-19},
 volume = {189},
 year = {1947}
}

@article{dsrep2,
 ISSN = {0003486X, 19398980},
 URL = {http://www.jstor.org/stable/1969376},
 author = {T. D. Newton},
 journal = {Annals of Mathematics},
 number = {3},
 pages = {730--733},
 publisher = {[Annals of Mathematics, Trustees of Princeton University on Behalf of the Annals of Mathematics, Mathematics Department, Princeton University]},
 title = {A Note on the Representations of the De Sitter Group},
 urldate = {2024-06-19},
 volume = {51},
 year = {1950}
}

@article{dsrep1,
 ISSN = {0003486X, 19398980},
 URL = {http://www.jstor.org/stable/1968990},
 author = {L. H. Thomas},
 journal = {Annals of Mathematics},
 number = {1},
 pages = {113--126},
 publisher = {[Annals of Mathematics, Trustees of Princeton University on Behalf of the Annals of Mathematics, Mathematics Department, Princeton University]},
 title = {On Unitary Representations of the Group of De Sitter Space},
 urldate = {2024-06-19},
 volume = {42},
 year = {1941}
}

@inproceedings{Ginsparg:1988ui,
    author = "Ginsparg, Paul H.",
    title = "{APPLIED CONFORMAL FIELD THEORY}",
    booktitle = "{Les Houches Summer School in Theoretical Physics: Fields, Strings, Critical Phenomena}",
    eprint = "hep-th/9108028",
    archivePrefix = "arXiv",
    reportNumber = "HUTP-88-A054",
    month = "9",
    year = "1988"
}

@article{Sun:2021thf,
    author = "Sun, Zimo",
    title = "{A note on the representations of SO(1,d + 1)}",
    eprint = "2111.04591",
    archivePrefix = "arXiv",
    primaryClass = "hep-th",
    doi = "10.1142/S0129055X24300073",
    journal = "Rev. Math. Phys.",
    volume = "37",
    number = "01",
    pages = "2430007",
    year = "2025"
}

@article{Bargmann:1946me,
    author = "Bargmann, V.",
    title = "{Irreducible unitary representations of the Lorentz group}",
    doi = "10.2307/1969129",
    journal = "Annals Math.",
    volume = "48",
    pages = "568--640",
    year = "1947"
}

@article{Penedones:2023uqc,
    author = "Penedones, Joao and Salehi Vaziri, Kamran and Sun, Zimo",
    title = "{Hilbert space of Quantum Field Theory in de Sitter spacetime}",
    eprint = "2301.04146",
    archivePrefix = "arXiv",
    primaryClass = "hep-th",
    month = "1",
    year = "2023"
}

@article{Bros:2010rku,
    author = "Bros, Jacques and Epstein, Henri and Moschella, Ugo",
    title = "{Particle decays and stability on the de Sitter universe}",
    eprint = "0812.3513",
    archivePrefix = "arXiv",
    primaryClass = "hep-th",
    doi = "10.1007/s00023-010-0042-7",
    journal = "Annales Henri Poincare",
    volume = "11",
    pages = "611--658",
    year = "2010"
}

@inproceedings{Flauger:2022hie,
    author = "Flauger, Raphael and Gorbenko, Victor and Joyce, Austin and McAllister, Liam and Shiu, Gary and Silverstein, Eva",
    title = "{Snowmass White Paper: Cosmology at the Theory Frontier}",
    booktitle = "{Snowmass 2021}",
    eprint = "2203.07629",
    archivePrefix = "arXiv",
    primaryClass = "hep-th",
    month = "3",
    year = "2022"
}

@article{Cohen:2020php,
    author = "Cohen, Timothy and Green, Daniel",
    title = "{Soft de Sitter Effective Theory}",
    eprint = "2007.03693",
    archivePrefix = "arXiv",
    primaryClass = "hep-th",
    doi = "10.1007/JHEP12(2020)041",
    journal = "JHEP",
    volume = "12",
    pages = "041",
    year = "2020"
}

@article{Cohen:2024anu,
    author = "Cohen, Timothy and Green, Daniel and Huang, Yiwen",
    title = "{Operator origin of anomalous dimensions in de Sitter space}",
    eprint = "2407.08581",
    archivePrefix = "arXiv",
    primaryClass = "hep-th",
    reportNumber = "CERN-TH-2024-103",
    doi = "10.1103/PhysRevD.111.103513",
    journal = "Phys. Rev. D",
    volume = "111",
    number = "10",
    pages = "103513",
    year = "2025"
}

@article{Marolf:2010zp,
    author = "Marolf, Donald and Morrison, Ian A.",
    title = "{The IR stability of de Sitter: Loop corrections to scalar propagators}",
    eprint = "1006.0035",
    archivePrefix = "arXiv",
    primaryClass = "gr-qc",
    doi = "10.1103/PhysRevD.82.105032",
    journal = "Phys. Rev. D",
    volume = "82",
    pages = "105032",
    year = "2010"
}

@article{Loparco:2023rug,
    author = "Loparco, Manuel and Penedones, Joao and Salehi Vaziri, Kamran and Sun, Zimo",
    title = {{The K\"all\'en-Lehmann representation in de Sitter spacetime}},
    eprint = "2306.00090",
    archivePrefix = "arXiv",
    primaryClass = "hep-th",
    doi = "10.1007/JHEP12(2023)159",
    journal = "JHEP",
    volume = "12",
    pages = "159",
    year = "2023"
}

@article{Hogervorst:2021uvp,
    author = "Hogervorst, Matthijs and Penedones, Jo\~ao and Vaziri, Kamran Salehi",
    title = "{Towards the non-perturbative cosmological bootstrap}",
    eprint = "2107.13871",
    archivePrefix = "arXiv",
    primaryClass = "hep-th",
    doi = "10.1007/JHEP02(2023)162",
    journal = "JHEP",
    volume = "02",
    pages = "162",
    year = "2023"
}

@article{Weinberg:2005vy,
    author = "Weinberg, Steven",
    title = "{Quantum contributions to cosmological correlations}",
    eprint = "hep-th/0506236",
    archivePrefix = "arXiv",
    reportNumber = "UTTG-01-05",
    doi = "10.1103/PhysRevD.72.043514",
    journal = "Phys. Rev. D",
    volume = "72",
    pages = "043514",
    year = "2005"
}

@article{Finelli:2008zg,
    author = "Finelli, F. and Marozzi, G. and Starobinsky, A. A. and Vacca, G. P. and Venturi, G.",
    title = "{Generation of fluctuations during inflation: Comparison of stochastic and field-theoretic approaches}",
    eprint = "0808.1786",
    archivePrefix = "arXiv",
    primaryClass = "hep-th",
    doi = "10.1103/PhysRevD.79.044007",
    journal = "Phys. Rev. D",
    volume = "79",
    pages = "044007",
    year = "2009"
}

@article{Gorbenko:2019rza,
    author = "Gorbenko, Victor and Senatore, Leonardo",
    title = "{$\lambda \phi^4$ in dS}",
    eprint = "1911.00022",
    archivePrefix = "arXiv",
    primaryClass = "hep-th",
    month = "10",
    year = "2019"
}

@article{Vennin:2015hra,
    author = "Vennin, Vincent and Starobinsky, Alexei A.",
    title = "{Correlation Functions in Stochastic Inflation}",
    eprint = "1506.04732",
    archivePrefix = "arXiv",
    primaryClass = "hep-th",
    doi = "10.1140/epjc/s10052-015-3643-y",
    journal = "Eur. Phys. J. C",
    volume = "75",
    pages = "413",
    year = "2015"
}

@article{Bardeen:1980kt,
    author = "Bardeen, James M.",
    title = "{Gauge Invariant Cosmological Perturbations}",
    doi = "10.1103/PhysRevD.22.1882",
    journal = "Phys. Rev. D",
    volume = "22",
    pages = "1882--1905",
    year = "1980"
}

@article{Weinberg:2003sw,
    author = "Weinberg, Steven",
    title = "{Adiabatic modes in cosmology}",
    eprint = "astro-ph/0302326",
    archivePrefix = "arXiv",
    reportNumber = "UTTG-12-02",
    doi = "10.1103/PhysRevD.67.123504",
    journal = "Phys. Rev. D",
    volume = "67",
    pages = "123504",
    year = "2003"
}

@article{Starobinsky:1986fx,
    author = "Starobinsky, Alexei A.",
    title = "{STOCHASTIC DE SITTER (INFLATIONARY) STAGE IN THE EARLY UNIVERSE}",
    doi = "10.1007/3-540-16452-9_6",
    journal = "Lect. Notes Phys.",
    volume = "246",
    pages = "107--126",
    year = "1986"
}

@article{Starobinsky:1994bd,
    author = "Starobinsky, Alexei A. and Yokoyama, Junichi",
    title = "{Equilibrium state of a selfinteracting scalar field in the De Sitter background}",
    eprint = "astro-ph/9407016",
    archivePrefix = "arXiv",
    reportNumber = "YITP-U-94-12",
    doi = "10.1103/PhysRevD.50.6357",
    journal = "Phys. Rev. D",
    volume = "50",
    pages = "6357--6368",
    year = "1994"
}

@article{Chakraborty:2023eoq,
    author = "Chakraborty, Priyesh and Stout, John",
    title = "{Compact scalars at the cosmological collider}",
    eprint = "2311.09219",
    archivePrefix = "arXiv",
    primaryClass = "hep-th",
    doi = "10.1007/JHEP03(2024)149",
    journal = "JHEP",
    volume = "03",
    pages = "149",
    year = "2024"
}

@article{Chakraborty:2023qbp,
    author = "Chakraborty, Priyesh and Stout, John",
    title = "{Light scalars at the cosmological collider}",
    eprint = "2310.01494",
    archivePrefix = "arXiv",
    primaryClass = "hep-th",
    doi = "10.1007/JHEP02(2024)021",
    journal = "JHEP",
    volume = "02",
    pages = "021",
    year = "2024"
}

@article{DiPietro:2021sjt,
    author = "Di Pietro, Lorenzo and Gorbenko, Victor and Komatsu, Shota",
    title = "{Analyticity and unitarity for cosmological correlators}",
    eprint = "2108.01695",
    archivePrefix = "arXiv",
    primaryClass = "hep-th",
    reportNumber = "CERN-TH-2021-118",
    doi = "10.1007/JHEP03(2022)023",
    journal = "JHEP",
    volume = "03",
    pages = "023",
    year = "2022"
}

@article{Bzowski:2015pba,
    author = "Bzowski, Adam and McFadden, Paul and Skenderis, Kostas",
    title = "{Scalar 3-point functions in CFT: renormalisation, beta functions and anomalies}",
    eprint = "1510.08442",
    archivePrefix = "arXiv",
    primaryClass = "hep-th",
    reportNumber = "IMPERIAL-TP-2015-PM-01",
    doi = "10.1007/JHEP03(2016)066",
    journal = "JHEP",
    volume = "03",
    pages = "066",
    year = "2016"
}

@article{Green:2020txs,
    author = "Green, Daniel and Premkumar, Akhil",
    title = "{Dynamical RG and Critical Phenomena in de Sitter Space}",
    eprint = "2001.05974",
    archivePrefix = "arXiv",
    primaryClass = "hep-th",
    doi = "10.1007/JHEP04(2020)064",
    journal = "JHEP",
    volume = "04",
    pages = "064",
    year = "2020"
}

@article{Cohen:2021jbo,
    author = "Cohen, Timothy and Green, Daniel and Premkumar, Akhil",
    title = "{A tail of eternal inflation}",
    eprint = "2111.09332",
    archivePrefix = "arXiv",
    primaryClass = "hep-th",
    doi = "10.21468/SciPostPhys.14.5.109",
    journal = "SciPost Phys.",
    volume = "14",
    number = "5",
    pages = "109",
    year = "2023"
}

@article{Cohen:2021fzf,
    author = "Cohen, Timothy and Green, Daniel and Premkumar, Akhil and Ridgway, Alexander",
    title = "{Stochastic Inflation at NNLO}",
    eprint = "2106.09728",
    archivePrefix = "arXiv",
    primaryClass = "hep-th",
    doi = "10.1007/JHEP09(2021)159",
    journal = "JHEP",
    volume = "09",
    pages = "159",
    year = "2021"
}

@article{Chakraborty:2025myb,
    author = "Chakraborty, Priyesh",
    title = "{Primordial Non-Gaussianity from Light Compact Scalars}",
    eprint = "2501.07672",
    archivePrefix = "arXiv",
    primaryClass = "hep-th",
    month = "1",
    year = "2025"
}

@article{Higuchi:1986py,
    author = "Higuchi, Atsushi",
    title = "{Forbidden Mass Range for Spin-2 Field Theory in De Sitter Space-time}",
    reportNumber = "YTP-86-06",
    doi = "10.1016/0550-3213(87)90691-2",
    journal = "Nucl. Phys. B",
    volume = "282",
    pages = "397--436",
    year = "1987"
}

@article{Suyama:2007bg,
    author = "Suyama, Teruaki and Yamaguchi, Masahide",
    title = "{Non-Gaussianity in the modulated reheating scenario}",
    eprint = "0709.2545",
    archivePrefix = "arXiv",
    primaryClass = "astro-ph",
    doi = "10.1103/PhysRevD.77.023505",
    journal = "Phys. Rev. D",
    volume = "77",
    pages = "023505",
    year = "2008"
}

@article{Smith:2011if,
    author = "Smith, Kendrick M. and LoVerde, Marilena and Zaldarriaga, Matias",
    title = "{A universal bound on N-point correlations from inflation}",
    eprint = "1108.1805",
    archivePrefix = "arXiv",
    primaryClass = "astro-ph.CO",
    doi = "10.1103/PhysRevLett.107.191301",
    journal = "Phys. Rev. Lett.",
    volume = "107",
    pages = "191301",
    year = "2011"
}

@article{Baumann:2022jpr,
    author = "Baumann, Daniel and Green, Daniel and Joyce, Austin and Pajer, Enrico and Pimentel, Guilherme L. and Sleight, Charlotte and Taronna, Massimo",
    title = "{Snowmass White Paper: The Cosmological Bootstrap}",
    eprint = "2203.08121",
    archivePrefix = "arXiv",
    primaryClass = "hep-th",
    doi = "10.21468/SciPostPhysCommRep.1",
    journal = "SciPost Phys. Comm. Rep.",
    volume = "2024",
    pages = "1",
    year = "2024"
}

@article{Achucarro:2022qrl,
    author = "Ach\'ucarro, Ana and others",
    title = "{Inflation: Theory and Observations}",
    eprint = "2203.08128",
    archivePrefix = "arXiv",
    primaryClass = "astro-ph.CO",
    month = "3",
    year = "2022"
}

@article{Maldacena:2002vr,
    author = "Maldacena, Juan Martin",
    title = "{Non-Gaussian features of primordial fluctuations in single field inflationary models}",
    eprint = "astro-ph/0210603",
    archivePrefix = "arXiv",
    doi = "10.1088/1126-6708/2003/05/013",
    journal = "JHEP",
    volume = "05",
    pages = "013",
    year = "2003"
}

@article{Creminelli:2004yq,
    author = "Creminelli, Paolo and Zaldarriaga, Matias",
    title = "{Single field consistency relation for the 3-point function}",
    eprint = "astro-ph/0407059",
    archivePrefix = "arXiv",
    reportNumber = "HUTP-04-A032",
    doi = "10.1088/1475-7516/2004/10/006",
    journal = "JCAP",
    volume = "10",
    pages = "006",
    year = "2004"
}

@article{Salopek:1990jq,
    author = "Salopek, D. S. and Bond, J. R.",
    title = "{Nonlinear evolution of long wavelength metric fluctuations in inflationary models}",
    reportNumber = "FERMILAB-PUB-90-131-A",
    doi = "10.1103/PhysRevD.42.3936",
    journal = "Phys. Rev. D",
    volume = "42",
    pages = "3936--3962",
    year = "1990"
}

@article{Green:2023uyz,
    author = "Green, Daniel and Guo, Yi and Han, Jiashu and Wallisch, Benjamin",
    title = "{Light fields during inflation from BOSS and future galaxy surveys}",
    eprint = "2311.04882",
    archivePrefix = "arXiv",
    primaryClass = "astro-ph.CO",
    doi = "10.1088/1475-7516/2024/05/090",
    journal = "JCAP",
    volume = "05",
    pages = "090",
    year = "2024"
}

@article{Cabass:2024wob,
    author = "Cabass, Giovanni and Philcox, Oliver H. E. and Ivanov, Mikhail M. and Akitsu, Kazuyuki and Chen, Shi-Fan and Simonovi\'c, Marko and Zaldarriaga, Matias",
    title = "{BOSS constraints on massive particles during inflation: The cosmological collider in action}",
    eprint = "2404.01894",
    archivePrefix = "arXiv",
    primaryClass = "astro-ph.CO",
    reportNumber = "RBI-ThPhys-2024-21, MIT-CTP/5698",
    doi = "10.1103/PhysRevD.111.063510",
    journal = "Phys. Rev. D",
    volume = "111",
    number = "6",
    pages = "063510",
    year = "2025"
}

@article{Philcox:2024jpd,
    author = "Philcox, Oliver H. E. and Kumar, Soubhik and Hill, J. Colin",
    title = "{Searching for inflationary particle production in Planck data}",
    eprint = "2405.03738",
    archivePrefix = "arXiv",
    primaryClass = "astro-ph.CO",
    doi = "10.1103/PhysRevD.111.103523",
    journal = "Phys. Rev. D",
    volume = "111",
    number = "10",
    pages = "103523",
    year = "2025"
}

@article{Assassi:2012zq,
    author = "Assassi, Valentin and Baumann, Daniel and Green, Daniel",
    title = "{On Soft Limits of Inflationary Correlation Functions}",
    eprint = "1204.4207",
    archivePrefix = "arXiv",
    primaryClass = "hep-th",
    doi = "10.1088/1475-7516/2012/11/047",
    journal = "JCAP",
    volume = "11",
    pages = "047",
    year = "2012"
}

@article{Pham:1985cr,
    author = "Pham, T. N. and Truong, Tran N.",
    title = "{Evaluation of the Derivative Quartic Terms of the Meson Chiral Lagrangian From Forward Dispersion Relation}",
    reportNumber = "Print-85-0588 (ECOLE POLY)",
    doi = "10.1103/PhysRevD.31.3027",
    journal = "Phys. Rev. D",
    volume = "31",
    pages = "3027",
    year = "1985"
}

@inproceedings{deRham:2022hpx,
    author = "de Rham, Claudia and Kundu, Sandipan and Reece, Matthew and Tolley, Andrew J. and Zhou, Shuang-Yong",
    title = "{Snowmass White Paper: UV Constraints on IR Physics}",
    booktitle = "{Snowmass 2021}",
    eprint = "2203.06805",
    archivePrefix = "arXiv",
    primaryClass = "hep-th",
    month = "3",
    year = "2022"
}

@article{Creminelli:2022onn,
    author = "Creminelli, Paolo and Janssen, Oliver and Senatore, Leonardo",
    title = "{Positivity bounds on effective field theories with spontaneously broken Lorentz invariance}",
    eprint = "2207.14224",
    archivePrefix = "arXiv",
    primaryClass = "hep-th",
    doi = "10.1007/JHEP09(2022)201",
    journal = "JHEP",
    volume = "09",
    pages = "201",
    year = "2022"
}

@article{Grall:2021xxm,
    author = "Grall, Tanguy and Melville, Scott",
    title = "{Positivity bounds without boosts: New constraints on low energy effective field theories from the UV}",
    eprint = "2102.05683",
    archivePrefix = "arXiv",
    primaryClass = "hep-th",
    doi = "10.1103/PhysRevD.105.L121301",
    journal = "Phys. Rev. D",
    volume = "105",
    number = "12",
    pages = "L121301",
    year = "2022"
}

@article{Baumann:2015nta,
    author = "Baumann, Daniel and Green, Daniel and Lee, Hayden and Porto, Rafael A.",
    title = "{Signs of Analyticity in Single-Field Inflation}",
    eprint = "1502.07304",
    archivePrefix = "arXiv",
    primaryClass = "hep-th",
    reportNumber = "ICTP-SAIFR-15-252",
    doi = "10.1103/PhysRevD.93.023523",
    journal = "Phys. Rev. D",
    volume = "93",
    number = "2",
    pages = "023523",
    year = "2016"
}

@article{Baumann:2019ghk,
    author = "Baumann, Daniel and Green, Daniel and Hartman, Thomas",
    title = "{Dynamical Constraints on RG Flows and Cosmology}",
    eprint = "1906.10226",
    archivePrefix = "arXiv",
    primaryClass = "hep-th",
    doi = "10.1007/JHEP12(2019)134",
    journal = "JHEP",
    volume = "12",
    pages = "134",
    year = "2019"
}

@article{Adams:2006sv,
    author = "Adams, Allan and Arkani-Hamed, Nima and Dubovsky, Sergei and Nicolis, Alberto and Rattazzi, Riccardo",
    title = "{Causality, analyticity and an IR obstruction to UV completion}",
    eprint = "hep-th/0602178",
    archivePrefix = "arXiv",
    reportNumber = "CERN-PH-TH-2006-033, HUTP-06-A0005",
    doi = "10.1088/1126-6708/2006/10/014",
    journal = "JHEP",
    volume = "10",
    pages = "014",
    year = "2006"
}

@article{Achucarro:2010da,
    author = "Achucarro, Ana and Gong, Jinn-Ouk and Hardeman, Sjoerd and Palma, Gonzalo A. and Patil, Subodh P.",
    title = "{Features of heavy physics in the CMB power spectrum}",
    eprint = "1010.3693",
    archivePrefix = "arXiv",
    primaryClass = "hep-ph",
    reportNumber = "LPTENS-10-36, CPHT-RR-080.0910",
    doi = "10.1088/1475-7516/2011/01/030",
    journal = "JCAP",
    volume = "01",
    pages = "030",
    year = "2011"
}

@article{Noumi:2012vr,
    author = "Noumi, Toshifumi and Yamaguchi, Masahide and Yokoyama, Daisuke",
    title = "{Effective field theory approach to quasi-single field inflation and effects of heavy fields}",
    eprint = "1211.1624",
    archivePrefix = "arXiv",
    primaryClass = "hep-th",
    reportNumber = "UT-KOMABA-12-9, TIT-HEP-625",
    doi = "10.1007/JHEP06(2013)051",
    journal = "JHEP",
    volume = "06",
    pages = "051",
    year = "2013"
}

@article{Baumann:2011nk,
    author = "Baumann, Daniel and Green, Daniel",
    title = "{Signatures of Supersymmetry from the Early Universe}",
    eprint = "1109.0292",
    archivePrefix = "arXiv",
    primaryClass = "hep-th",
    doi = "10.1103/PhysRevD.85.103520",
    journal = "Phys. Rev. D",
    volume = "85",
    pages = "103520",
    year = "2012"
}

@article{Arkani-Hamed:2015bza,
    author = "Arkani-Hamed, Nima and Maldacena, Juan",
    title = "{Cosmological Collider Physics}",
    eprint = "1503.08043",
    archivePrefix = "arXiv",
    primaryClass = "hep-th",
    month = "3",
    year = "2015"
}

@article{Chen:2009zp,
    author = "Chen, Xingang and Wang, Yi",
    title = "{Quasi-Single Field Inflation and Non-Gaussianities}",
    eprint = "0911.3380",
    archivePrefix = "arXiv",
    primaryClass = "hep-th",
    doi = "10.1088/1475-7516/2010/04/027",
    journal = "JCAP",
    volume = "04",
    pages = "027",
    year = "2010"
}

@article{Mirbabayi:2015hva,
    author = "Mirbabayi, Mehrdad and Simonovi\'c, Marko",
    title = "{Effective Theory of Squeezed Correlation Functions}",
    eprint = "1507.04755",
    archivePrefix = "arXiv",
    primaryClass = "hep-th",
    doi = "10.1088/1475-7516/2016/03/056",
    journal = "JCAP",
    volume = "03",
    pages = "056",
    year = "2016"
}

@article{Philcox:2025wts,
    author = "Philcox, Oliver H. E.",
    title = "{Searching for Inflationary Physics with the CMB Trispectrum: 3. Constraints from Planck}",
    eprint = "2502.06931",
    archivePrefix = "arXiv",
    primaryClass = "astro-ph.CO",
    month = "2",
    year = "2025"
}

@article{Cabass:2022oap,
    author = "Cabass, Giovanni and Ivanov, Mikhail M. and Philcox, Oliver H. E.",
    title = "{Colliders and ghosts: Constraining inflation with the parity-odd galaxy four-point function}",
    eprint = "2210.16320",
    archivePrefix = "arXiv",
    primaryClass = "astro-ph.CO",
    doi = "10.1103/PhysRevD.107.023523",
    journal = "Phys. Rev. D",
    volume = "107",
    number = "2",
    pages = "023523",
    year = "2023"
}

@article{Slosar:2008hx,
    author = "Slosar, Anze and Hirata, Christopher and Seljak, Uros and Ho, Shirley and Padmanabhan, Nikhil",
    title = "{Constraints on local primordial non-Gaussianity from large scale structure}",
    eprint = "0805.3580",
    archivePrefix = "arXiv",
    primaryClass = "astro-ph",
    doi = "10.1088/1475-7516/2008/08/031",
    journal = "JCAP",
    volume = "08",
    pages = "031",
    year = "2008"
}

@article{Munchmeyer:2019wlh,
    author = {M\"unchmeyer, Moritz and Smith, Kendrick M.},
    title = "{Higher N-point function data analysis techniques for heavy particle production and WMAP results}",
    eprint = "1910.00596",
    archivePrefix = "arXiv",
    primaryClass = "astro-ph.CO",
    doi = "10.1103/PhysRevD.100.123511",
    journal = "Phys. Rev. D",
    volume = "100",
    number = "12",
    pages = "123511",
    year = "2019"
}

@inbook{Green:2022ovz,
    author = "Green, Daniel",
    title = "{EFT for de Sitter Space}",
    eprint = "2210.05820",
    archivePrefix = "arXiv",
    primaryClass = "hep-th",
    doi = "10.1007/978-981-19-3079-9_6-1",
    year = "2023"
}

@article{Tolley:2001gg,
    author = "Tolley, Andrew J. and Turok, Neil",
    title = "{Quantization of the massless minimally coupled scalar field and the dS / CFT correspondence}",
    eprint = "hep-th/0108119",
    archivePrefix = "arXiv",
    month = "8",
    year = "2001"
}

@article{Loparco:2025azm,
    author = "Loparco, Manuel and Penedones, Joao and Ulrich, Yannis",
    title = "{What is a photon in de Sitter spacetime?}",
    eprint = "2505.00761",
    archivePrefix = "arXiv",
    primaryClass = "hep-th",
    month = "5",
    year = "2025"
}

@article{Burgess:2015ajz,
    author = "Burgess, C. P. and Holman, R. and Tasinato, G.",
    title = "{Open EFTs, IR effects {\textbackslash}{\&} late-time resummations: systematic corrections in stochastic inflation}",
    eprint = "1512.00169",
    archivePrefix = "arXiv",
    primaryClass = "gr-qc",
    doi = "10.1007/JHEP01(2016)153",
    journal = "JHEP",
    volume = "01",
    pages = "153",
    year = "2016"
}

@article{Baumgart:2019clc,
    author = "Baumgart, Matthew and Sundrum, Raman",
    title = "{De Sitter Diagrammar and the Resummation of Time}",
    eprint = "1912.09502",
    archivePrefix = "arXiv",
    primaryClass = "hep-th",
    doi = "10.1007/JHEP07(2020)119",
    journal = "JHEP",
    volume = "07",
    pages = "119",
    year = "2020"
}

@article{Mirbabayi:2019qtx,
    author = "Mirbabayi, Mehrdad",
    title = "{Infrared dynamics of a light scalar field in de Sitter}",
    eprint = "1911.00564",
    archivePrefix = "arXiv",
    primaryClass = "hep-th",
    doi = "10.1088/1475-7516/2020/12/006",
    journal = "JCAP",
    volume = "12",
    pages = "006",
    year = "2020"
}

@article{Vilenkin:1983xq,
    author = "Vilenkin, Alexander",
    title = "{The Birth of Inflationary Universes}",
    reportNumber = "TUTP-83-1",
    doi = "10.1103/PhysRevD.27.2848",
    journal = "Phys. Rev. D",
    volume = "27",
    pages = "2848",
    year = "1983"
}

@article{Aryal:1987vn,
    author = "Aryal, Mukunda and Vilenkin, Alexander",
    title = "{The Fractal Dimension of Inflationary Universe}",
    reportNumber = "TUTP-87-11",
    doi = "10.1016/0370-2693(87)90932-4",
    journal = "Phys. Lett. B",
    volume = "199",
    pages = "351--357",
    year = "1987"
}

@article{Green:2013rd,
    author = "Green, Daniel and Lewandowski, Matthew and Senatore, Leonardo and Silverstein, Eva and Zaldarriaga, Matias",
    title = "{Anomalous Dimensions and Non-Gaussianity}",
    eprint = "1301.2630",
    archivePrefix = "arXiv",
    primaryClass = "hep-th",
    reportNumber = "SLAC-PUB-15334, SU-ITP-12-42",
    doi = "10.1007/JHEP10(2013)171",
    journal = "JHEP",
    volume = "10",
    pages = "171",
    year = "2013"
}

@article{Mirbabayi:2020vyt,
    author = "Mirbabayi, Mehrdad",
    title = "{Markovian dynamics in de Sitter}",
    eprint = "2010.06604",
    archivePrefix = "arXiv",
    primaryClass = "hep-th",
    doi = "10.1088/1475-7516/2021/09/038",
    journal = "JCAP",
    volume = "09",
    pages = "038",
    year = "2021"
}

@article{Bros:1990cu,
    author = "Bros, Jacques",
    title = "{Complexified de Sitter space: Analytic causal kernels and Kallen-Lehmann type representation}",
    reportNumber = "SACLAY-SPH-T-90-114",
    doi = "10.1016/0920-5632(91)90119-Y",
    journal = "Nucl. Phys. B Proc. Suppl.",
    volume = "18",
    pages = "22--28",
    year = "1991"
}

@article{Lu:2021wxu,
    author = "Lu, Qianshu and Reece, Matthew and Xianyu, Zhong-Zhi",
    title = "{Missing scalars at the cosmological collider}",
    eprint = "2108.11385",
    archivePrefix = "arXiv",
    primaryClass = "hep-ph",
    doi = "10.1007/JHEP12(2021)098",
    journal = "JHEP",
    volume = "12",
    pages = "098",
    year = "2021"
}

@article{Fitzpatrick:2011hu,
    author = "Fitzpatrick, A. Liam and Kaplan, Jared",
    title = "{Analyticity and the Holographic S-Matrix}",
    eprint = "1111.6972",
    archivePrefix = "arXiv",
    primaryClass = "hep-th",
    reportNumber = "SLAC-PUB-14841",
    doi = "10.1007/JHEP10(2012)127",
    journal = "JHEP",
    volume = "10",
    pages = "127",
    year = "2012"
}

@article{Giombi:2017hpr,
    author = "Giombi, Simone and Sleight, Charlotte and Taronna, Massimo",
    title = "{Spinning AdS Loop Diagrams: Two Point Functions}",
    eprint = "1708.08404",
    archivePrefix = "arXiv",
    primaryClass = "hep-th",
    reportNumber = "PUPT-2540",
    doi = "10.1007/JHEP06(2018)030",
    journal = "JHEP",
    volume = "06",
    pages = "030",
    year = "2018"
}

@article{Reece:2025thc,
    author = "Reece, Matthew",
    title = "{Extra-dimensional axion expectations}",
    eprint = "2406.08543",
    archivePrefix = "arXiv",
    primaryClass = "hep-ph",
    doi = "10.1007/JHEP07(2025)130",
    journal = "JHEP",
    volume = "07",
    pages = "130",
    year = "2025"
}

@article{SalehiVaziri:2024joi,
    author = "Salehi Vaziri, Kamran",
    title = "{A non-perturbative construction of the de Sitter late-time boundary}",
    eprint = "2412.00183",
    archivePrefix = "arXiv",
    primaryClass = "hep-th",
    month = "11",
    year = "2024"
}

@book{Weinberg:1995mt,
    author = "Weinberg, Steven",
    title = "{The Quantum theory of fields. Vol. 1: Foundations}",
    doi = "10.1017/CBO9781139644167",
    isbn = "978-0-521-67053-1, 978-0-511-25204-4",
    publisher = "Cambridge University Press",
    month = "6",
    year = "2005"
}

@article{Burgess:2009bs,
    author = "Burgess, C. P. and Leblond, L. and Holman, R. and Shandera, S.",
    title = "{Super-Hubble de Sitter Fluctuations and the Dynamical RG}",
    eprint = "0912.1608",
    archivePrefix = "arXiv",
    primaryClass = "hep-th",
    doi = "10.1088/1475-7516/2010/03/033",
    journal = "JCAP",
    volume = "03",
    pages = "033",
    year = "2010"
}

@article{Prokopec:2017vxx,
    author = "Prokopec, Tomislav and Rigopoulos, Gerasimos",
    title = "{Functional renormalization group for stochastic inflation}",
    eprint = "1710.07333",
    archivePrefix = "arXiv",
    primaryClass = "gr-qc",
    doi = "10.1088/1475-7516/2018/08/013",
    journal = "JCAP",
    volume = "08",
    pages = "013",
    year = "2018"
}

@article{Cespedes:2023aal,
    author = "C{\'e}spedes, Sebasti{\'a}n and Davis, Anne-Christine and Wang, Dong-Gang",
    title = "{On the IR divergences in de Sitter space: loops, resummation and the semi-classical wavefunction}",
    eprint = "2311.17990",
    archivePrefix = "arXiv",
    primaryClass = "hep-th",
    doi = "10.1007/JHEP04(2024)004",
    journal = "JHEP",
    volume = "04",
    pages = "004",
    year = "2024"
}

\end{document}